\documentclass[journal]{IEEEtran}
\usepackage[export]{adjustbox}
\usepackage[linesnumbered,ruled,vlined]{algorithm2e}
\usepackage{graphicx}

\usepackage{boxedminipage}
\usepackage{caption}

\usepackage{listings}

\usepackage{times}
\usepackage[cmex10]{amsmath}
\usepackage{amssymb,amsthm,comma}
\usepackage{tabularx}
\usepackage{hhline}
\usepackage{hyperref}
\usepackage{subcaption}
\usepackage{url}
\usepackage[table,xcdraw]{xcolor}
\usepackage{xspace}
\usepackage{booktabs}
\usepackage{multirow}
\usepackage{mathtools}

\newtheorem{example}{Example}
\newtheorem{definition}{Definition}
\hypersetup{
    colorlinks,
    linkcolor={red!50!black},
    citecolor={blue!50!black},
    urlcolor={blue!80!black}
}
\newcommand\step[1]{\item \textbf{Step} #1:}

\newcommand{\edit}[1]{ {#1}}
\newcommand{\ccol}{\cellcolor[HTML]{AFEFAF}}
\newcommand{\comp}{\cellcolor[HTML]{FFA365}}
\newcommand{\res}{{\cellcolor[HTML]{5CC5EB}}}  
\SetKwRepeat{Do}{do}{while} 
\theoremstyle{exampstyle} \newtheorem{theorem}{Theorem}[section]
\theoremstyle{exampstyle}

\newcommand\blfootnote[1]{%
  \begingroup
  \renewcommand\thefootnote{}\footnote{#1}%
  \addtocounter{footnote}{-1}%
  \endgroup
}
\begin{document}

\title{Crossbar-Constrained Technology Mapping for ReRAM based In-Memory Computing}

\author{\IEEEauthorblockN{Debjyoti Bhattacharjee, Yaswanth Tavva, Arvind Easwaran, Anupam Chattopadhyay}\\
	\IEEEauthorblockA{ School of Computer Science and Engineering,\\
	Nanyang Technological University, Singapore - 639798\\
		Email:  \it{\{debjyoti001,yaswanth001,arvinde,anupam\}@ntu.edu.sg}}}

\maketitle

 \begin{abstract}
In recent times, Resistive RAMs (ReRAMs) have gained significant prominence due to their unique feature of supporting both non-volatile storage and logic capabilities. ReRAM is also reported to provide extremely low power consumption compared to the standard CMOS storage devices. As a result, researchers have explored the mapping and design of diverse applications, ranging from arithmetic to neuromorphic computing structures to ReRAM-based platforms. ReVAMP, a general-purpose ReRAM computing platform, has been proposed recently to leverage the parallelism exhibited in a crossbar structure. However, the technology mapping on ReVAMP remains an open challenge. Though the technology mapping with device/area-constraints have been proposed, crossbar constraints are not considered so far. In this work, we address this problem. Two technology mapping flows are proposed, considering different runtime-efficiency trade-offs. Both the mapping flows take crossbar constraints into account and generate feasible mapping for a variety of crossbar dimensions. Our proposed algorithms are highly scalable and reveal important design hints for ReRAM-based implementations.
\end{abstract}

 \blfootnote{This work is an extension of the following publication. Bhattacharjee D, Devadoss R, Chattopadhyay A. ReVAMP: ReRAM based VLIW architecture for in-memory computing. In2017 Design, Automation \& Test in Europe Conference \& Exhibition (DATE) 2017 Mar 27 (pp. 782-787). IEEE.}
 \section{Introduction}\label{sec:intro}
\noindent { Traditional computing platforms require transfer of data along energy-hungry buses between the compute cores and the memory hierarchy. This has resulted in performance degradation~(memory wall) leading to challenges while processing big-data~\cite{babarinsa2015jafar,koo2017summarizer}.
Data transfer between cores and memory is often costlier than computing itself~\cite{ahn2015pim}. Such challenges can be mitigated by logic-in-memory~(LiM) enabled devices, which can perform simple Boolean operations within the memory or very close to the memory itself. \edit{Efficient algorithms for LiM can lead to considerable improvements in performance of applications that require large memory bandwidth to process inputs~\cite{ahn2016scalable, hsieh2016transparent, seshadri2017ambit,kim2017grim}.} Therefore, we need to build mapping tools to leverage the benefits of LiM architectures.}

One of the most promising emerging non-volatile memory with computation capabilities is Resistive RAM~(ReRAM). ReRAMs offer fast read/write speeds~\cite{torrezan2011sub}, high endurance~\cite{lee2011fast}, long retention times~\cite{wei2012retention} along with the scope of 3D fabrication~\cite{chien2012multi}.  Large passive crossbar  arrays  can  be  enabled by preventing parasitic currents  by  means  of  devices  such  as  a  select device  in  series  to  a  switch~(1S1R)  or  a  Complementary  Resistive Switch~(CRS)~\cite{linn2010complementary}. Unlike CRS devices,  1S1R devices offer non-destructive read outs making them suitable for logic in memory operations. ReRAMs are fast gaining popularity for use as computation devices.  Recently, multiple propositions for realizing 
arithmetic blocks using ReRAMs have been proposed~\cite{siemon2015complementary,biblas}. In addition, efficient implementations of encryption, data compression and linear algebra algorithms have also been mapped to ReRAMs~\cite{bhattacharjee2017sha,bhattacharjee2017memory,xia2016technological}. { ReRAMs have been also used for neuromorphic computation~\cite{hycm}.
\edit{Analog non-volatile ReRAM based synapses have been used 
for gray-scale face classification for energy savings~\cite{yao2017face}. Even, emulation of metaplasticity has been demonstrated using analog ReRAMs~\cite{zhu2017emulation}. Analog memristor crossbar arrays have also been used for sparse encoding of input data, which can be extended for image processing applications~\cite{sheridan2017sparse}. 
Further, to enable uniform analog switching, fast speed, along with excellent retention properties, a thermal enhanced layer has been proposed to confine heat in switching layer~\cite{wu2017improving}. Multi-state memristors have also been used for ternary arithmetic as well as native multi-valued logic implementation~\cite{kim2016multistate,bhattacharjee2018multi}. Ot has been experimentally demonstrated that 3D-fabrication is feasible for resistive RAM arrays~\cite{bai2014study}.}

From the perspective of computing arbitrary Boolean functions, a preliminary method for computing using memristors realizing material implication, was presented by Lehtonen et al.~\cite{lehtonen2009stateful}. Further, it was shown that any arbitrary Boolean expression can be computed using two working memristors that realize material implication~\cite{lehtonen2010two}.  Logic synthesis flows have been proposed using Imply Sequence
Diagram and Or-Invertor Graph for memristors realizing material implication~\cite{Raghuvanshi2014,oiganu}. Optimal technology mapping for ReRAM devices have been investigated for ReRAM devices, that realize three-input Boolean majority with a single input inverted~\cite{bhattacharjee2016delay}. In addition, area-constrained technology mapping for individual ReRAM devices using Integer Linear Programming, along with scalable heuristics have been proposed~\cite{aomap}.  A general purpose bit-serial Programmable Logic in Memory~(PLiM) architecture was proposed~\cite{Gaillardon} that uses ReRAM crossbar for data storage as well as computation. {\edit A compiler for the same was developed by Soeken et al.~\cite{soeken2016mig}.} However, these works either consider independent devices or use serial operations on ReRAM crossbar arrays. A transpose resistive memory with additional controller circuitry, was proposed by Nishil et al.~\cite{talati2016logic}, for which a technology mapping was proposed recently~\cite{hursimple}. Inherently, ReRAM arrays support operations on multiple devices that are on the same wordline, allowing parallel operations. The ReVAMP architecture allows harnessing this parallelism by means of VLIW instructions~\cite{bhattacharjee2017revamp}.

A ReRAM crossbar array consists of multiple ReRAM devices that share wordlines and bitlines.
In this paper, we address the problem 
of technology mapping for computation using ReRAM crossbar array, by using ReVAMP as the target logic-in-memory architecture. The main challenge is to efficiently harness the bit-level parallelism offered 
by the crossbar arrays. The key contributions of the paper are as follows.
\begin{itemize}
 \item Any arbitrary AIG/MIG with $k$-levels can be mapped with $2(k+1)$ devices, arranged as a crossbar with at least two bitlines. 
 \item Any Boolean expression, expressed as a Exclusive-Sum-Of-Product~(ESOP), can be computed on a crossbar with three wordlines and at least two bitlines. 
 \item We present two technology mapping approaches for ReVAMP in-memory computing platform.
 \item The area-constrained technology mapping approach uses And-Inverter Graph for logic representation and then uses a hierarchical method for generating ReVAMP instructions, aware of
 the crossbar dimensions. The method supports mapping to a wide variety of crossbar dimensions.
 \item The delay-constrained mapping approach relies on harnessing bit-level parallelism of the ReRAM crossbar array by maximizing parallel operations across multiple devices that share the same wordline. This method achieves significant lower delay compared to existing ReRAM-based serial logic-in-memory architecture.
\end{itemize}

The rest of the paper is organized as follows. In section~\ref{sec:prelim}, we present an introduction to ReVAMP, along with a brief introduction to Boolean logic networks.
Section~\ref{sec:problem} formally presents the technology mapping problem followed by outline of the solution approaches. Section~\ref{sec:area} describes the solution for the area-constrained
technology mapping. Section~\ref{sec:delay} presents a technology mapping solution for fast mapping by exploiting inherent crossbar parallelism. Benchmarking results are presented in
section~\ref{sec:experimental}. Section~\ref{sec:conc} concludes the paper.
 \section{Preliminaries}\label{sec:prelim}
\noindent In this section, we present the details of logic operations using ReRAM crossbar arrays, followed 
by ReVAMP --- a ReRAM based general purpose computing architecture. We also summarily present the details of Boolean logic networks which will be used for technology mapping.

\subsection{Logic in memory operations using ReRAM crossbar arrays}
\noindent The ReRAM device model proposed in~\cite{siemon2014simulation}, was fitted to a $Pt/(11 nm)TaO_x/Ta$ cell. The used selector device is the $Pt/TaO_x/TiO_2/TaO_x/Pt$ crested barrier device proposed in~\cite{kim2012selector, lee2012high}. Both devices were implemented in VerilogA and simulated using Cadence Spectre. The used ReRAM model considers a filamentary region in which the switching takes place by a redistribution of ionic defects, i.e., oxygen vacancies. The filament is modeled by three lumped circuit elements: a Schottky-type diode representing the current flow through the $Pt/TaO_x$ interface, a disc resistance describing the region close to the Schottky-type interface and a resistance, which comprises the plug resistance describing the remaining part of the filament and the resistance of the electrodes. The state variable of the resistive switching model is the oxygen vacancy concentration $N$ close to the active electrode interface, which modulates the disc resistance and the electron transport through the Schottky-type diode. 

 For logic operations, each ReRAM device can be interpreted as a finite-state-machine~(FSM), as shown in Fig.~\ref{fig:device}. Each device has two input terminals---the wordline~$wl$ and the bitline~$bl$. 
The internal resistive state $Z$ of the ReRAM acts as a third input and the stored bit. If the state $Z$ is
in High Resistive State~(HRS), it is interpreted as logic 0, while Low Resistive State~(LRS) is interpreted as
logic 1.
As shown in following equation, the next state of the device~$Z_n$ is expressed as a $3$-input majority function, with the bitline input inverted. 
\begin{equation} \label{eq:m3}
 Z_n = M_3(Z,wl,\overline{bl})
\end{equation}
This forms the fundamental logic operation that can be realized using ReRAM devices. 
The inversion operation is equivalent to using the intrinsic function $Z_n$ with one input~(wordline or state) as 0, the second input~(state or wordline) at 1 and the bitline input as the variable to be inverted. 
\begin{equation} \label{eq:inv}
 \overline{v} = M_3(0,1,\overline{v}) = M_3(1,0,\overline{v}) 
\end{equation}
Since majority and inversion operations form a \textit{functionally complete} set, any Boolean function can be realized using only $Z_n$ operations.

\begin{figure}[t]
\begin{center}
\includegraphics[width=2.8in]{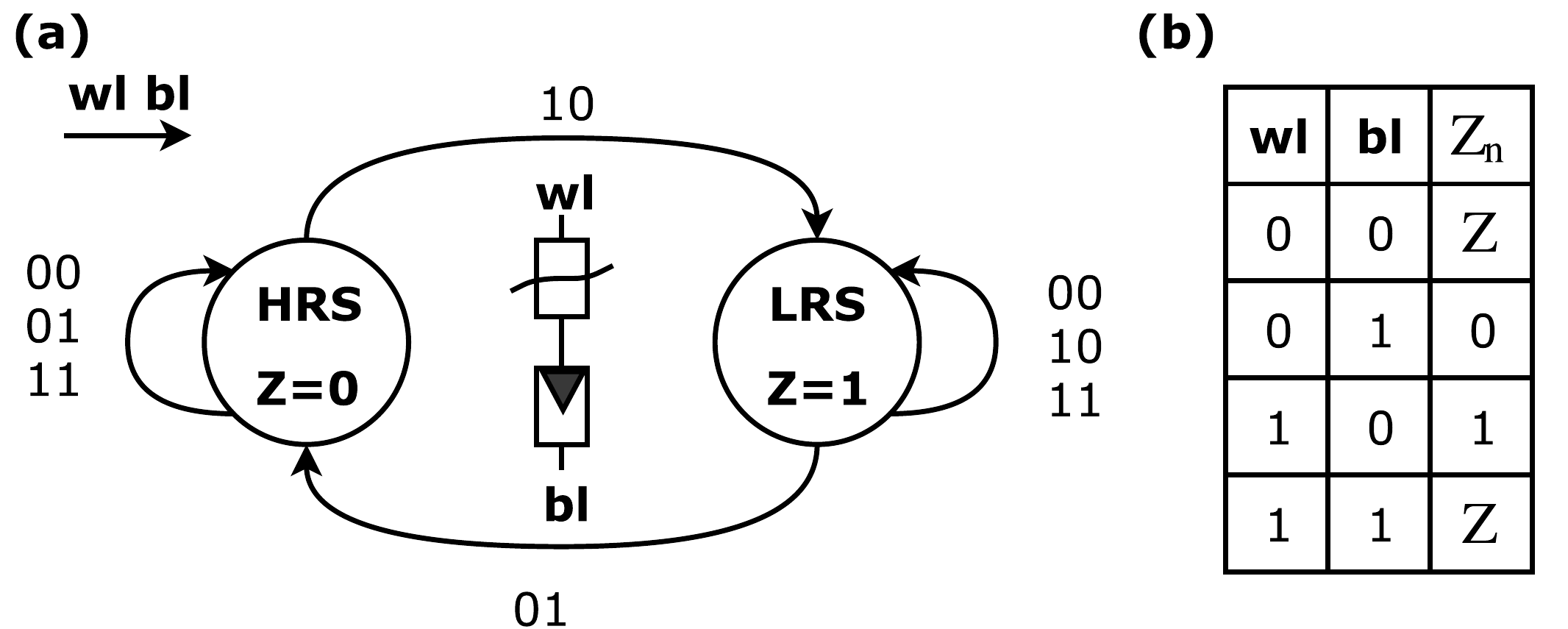}
\end{center}
\caption{\em Logic operation using 1S1R device. (a)~FSM. Wordline input 1 and bitline input 0  changes the device state to logic 1 whereas
wordline input 0 and bitline input 1  changes the device state to logic 0. Other input combinations do not change the internal device state $Z$.
(b)~Truth Table of the intrinsic function $Z_n$.}
\label{fig:device}
\vspace{-0.5cm}
\end{figure}

\begin{figure}[t]
\begin{center}
\includegraphics[width=2.8in]{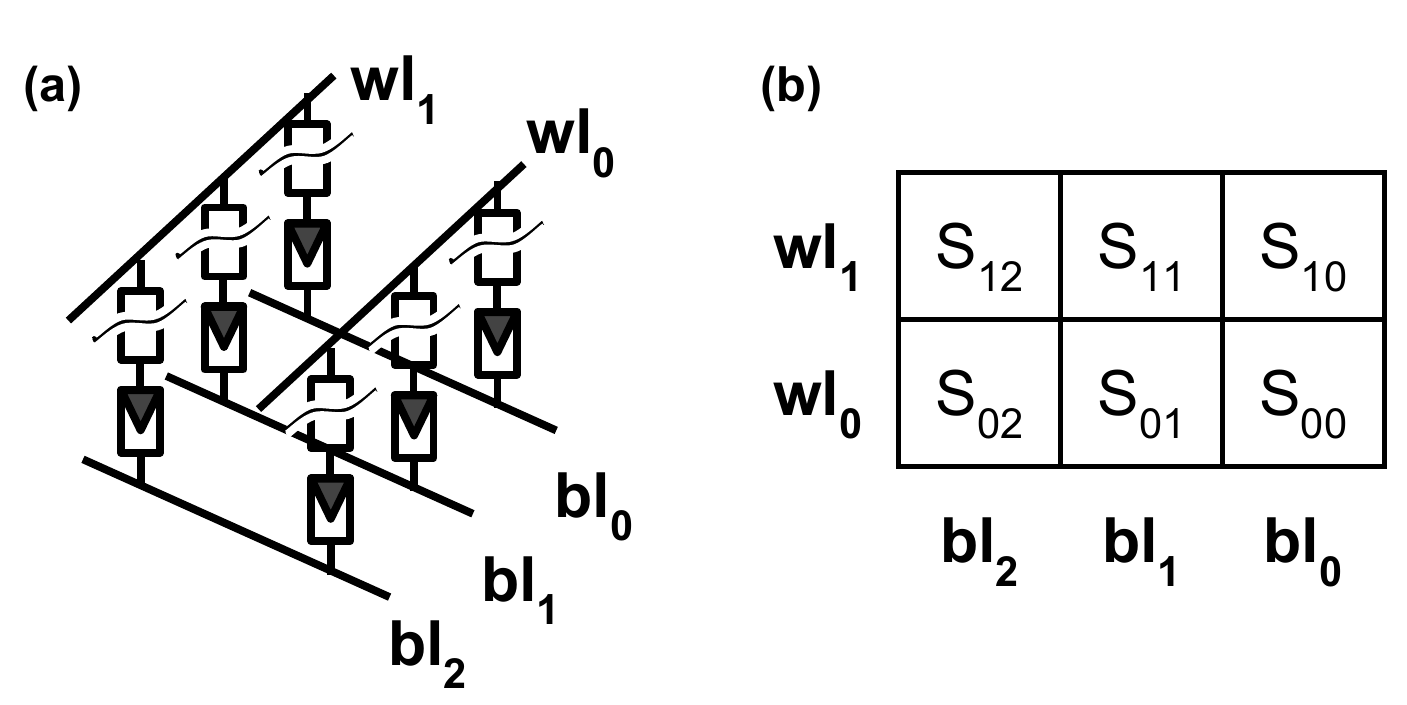}
\end{center}
\caption{\em A $2\times 3$ ReRAM crossbar array (a)~Six 1S1R devices arranged as crossbar. (b)~The crossbar represented as a schematic. $S_{ij}$ represents internal state of device at wordline $i$ and bitline~$j$.  $wl_{i}$ represents the $i^{th}$ wordline input while $bl_{j}$ represents the $j^{th}$ bitline input.}
\label{fig:crossbar}
\end{figure}
 \begin{figure}[t]
 \centering
 \includegraphics[width=0.8\linewidth]{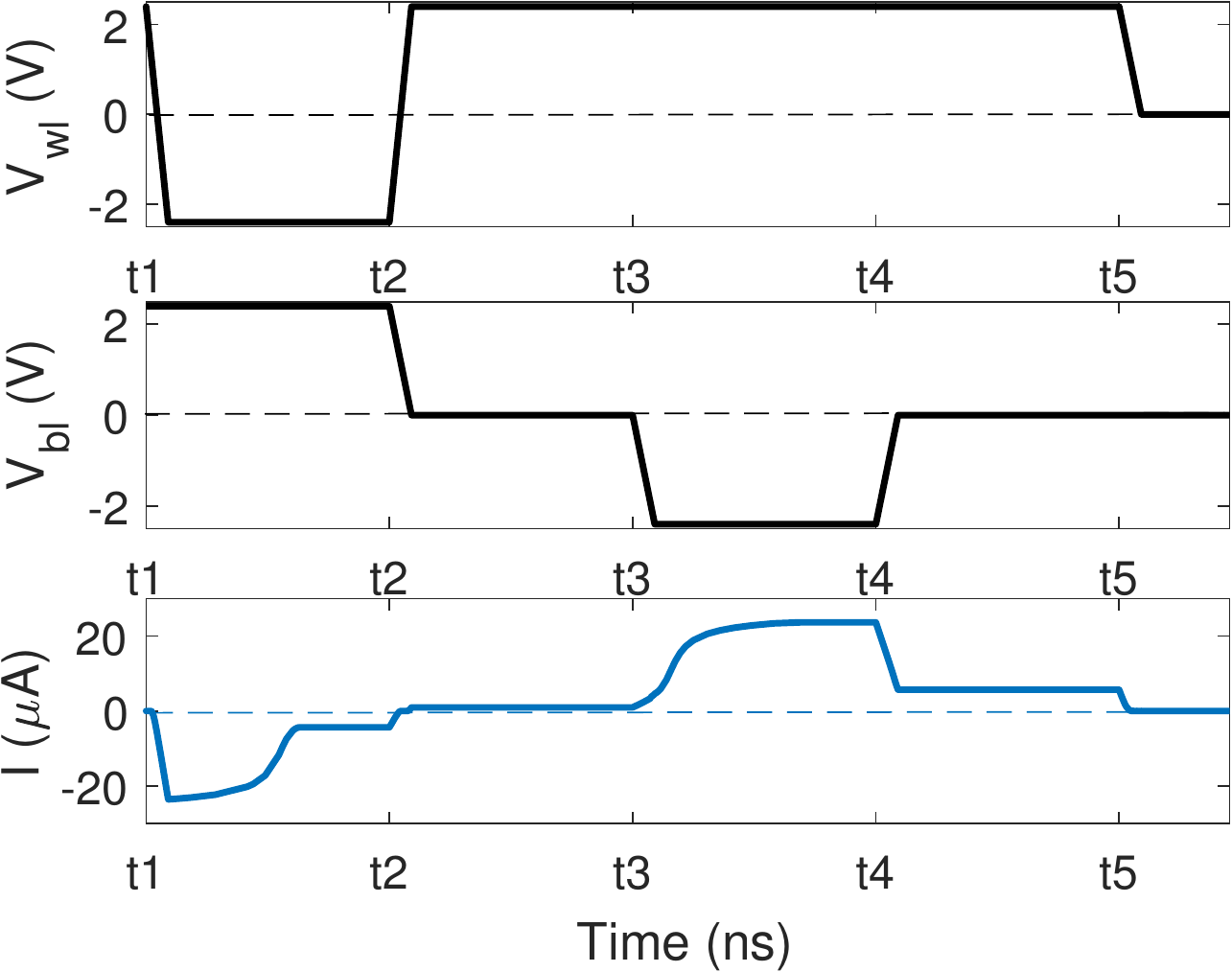}
 \caption{\em Cadence simulation of a 1S1R device. $V_{wl}$ and $V_{bl}$ indicate the applied wordline and bitline voltages. $I$ indicates the read out current, observed at the bitline.}
 \label{fig:wave}
 \vspace{-0.8cm}
\end{figure}

\begin{figure*}[!htbp]
\begin{center}
\includegraphics[width=5.8in]{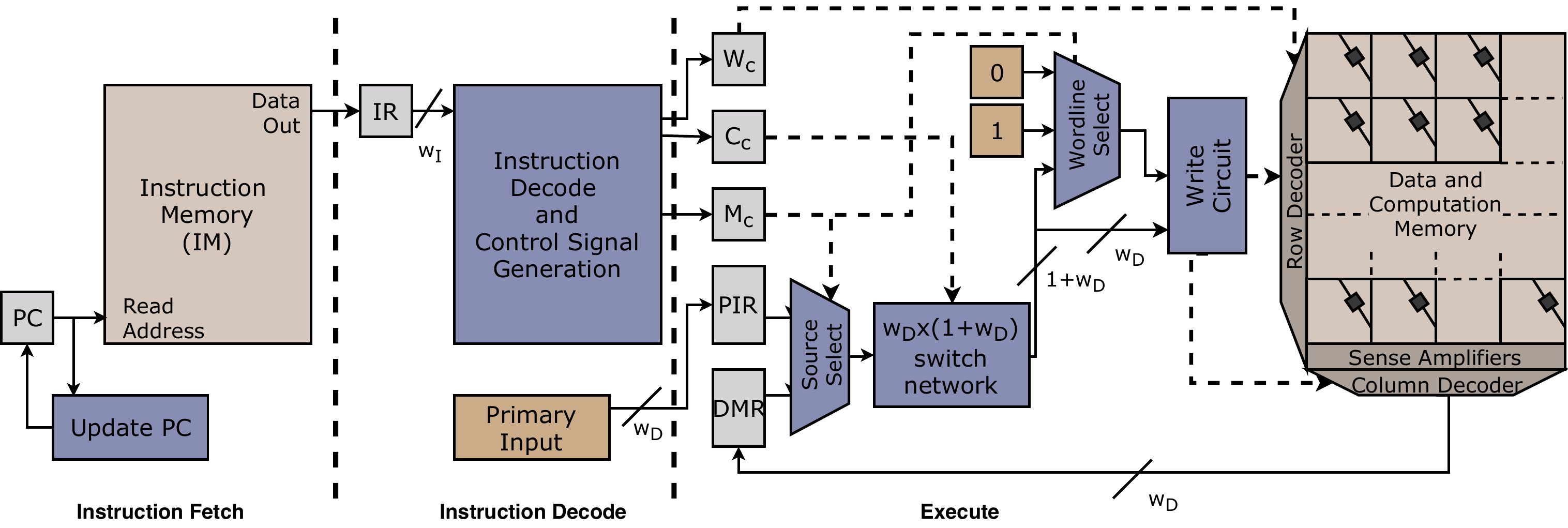}
\end{center}
\caption{\em ReVAMP architecture}
\label{fig:arch}
 \vspace{-0.8cm}
\end{figure*}
A ReRAM crossbar memory consists of multiple
1S1R ReRAM devices, arranged in the form of a crossbar array~\cite{eike_logic}. Multiple devices 
share wordlines and bitlines. Fig.~\ref{fig:crossbar} shows a ReRAM crossbar array with 6 devices arranged in $2\times3$ configuration i.e. 2 wordlines and 3 bitlines. The internal state of device $D_{ij}$ at wordline $i$ and bitline $j$ is referred as $S_{ij}$. The devices $D_{00}, D_{01}$ and $D_{02}$  share wordline 0 whereas the devices $D_{10}, D_{11}$ and $D_{12}$  share wordline 1.
Similarly, the devices $D_{00}$ and $D_{10}$ share bitline 0 and so on. 
 Like conventional RAM arrays, ReRAM memories are accessed as words. It should be noted that all the devices in a word share a common wordline. For example, 
 word 0 has devices $D_{00}, D_{01}$ and $D_{02}$.  
The ReRAM array is programmed using a V/2 scheme, with V$ =4.8V$. Logic 1 and 0 are realized by voltage pulses of $2.4V$ and $-2.4V$ respectively. Unselected lines are kept grounded. In a readout phase, the presence of a current greater than \mbox{$4 \mu A$}  implies logic~1 while its absence implies logic~0. 

Fig.~\ref{fig:wave} shows the Cadence simulation for a single device. In cycle $t1$,
0 and 1 are applied to the wordline and bitline, respectively to set the logic state to 0~(HRS). In cycle $t2$, the device state is read out. The read out current is less than $5\mu A$, confirming the device is in logic state 0. In the next cycle $t3$, 1 and 0 are applied to the wordline and bitline respectively, to set the logic state to 1~(LRS).  In $t4$, the devices is read out and the read out current is greater than $4\mu A$, indicating the logic state to be state 1. 

\subsection{ReVAMP architecture} \label{subsec:revamp}
\noindent We briefly present the \textbf{Re}RAM based \textbf{V}LIW \textbf{A}rchitecture for  in-\textbf{M}emory com\textbf{P}uting~(ReVAMP), depicted in Fig.~\ref{fig:arch}.
The architecture uses two ReRAM crossbar memories --- the Instruction Memory~(IM) and the Data Storage and Computation Memory~(DCM).
The IM is a regular instruction memory accessed using the program counter~(PC). 
The DCM hosts data and in-memory computation.
All the devices in one single word of the DCM is can be operated in parallel, with each operation being the intrinsic~$Z_n$ function. Since multiple~$Z_n$ operations operate in parallel, the proposed architecture is VLIW in nature. 
Splitting the instruction and data memory allows reduction in overall execution time, by pipelining instruction fetch and computation.  
The ReVAMP architecture is parameterized as shown in Table~\ref{table:param}, and can be configured as necessary. 

  \begin{table}[t]
  \centering
  \caption{ReVAMP parameters.}
  \label{table:param}
   \begin{tabular}{ll} \\ \hline
   \textbf{Parameter} & \textbf{Description} \\ \hline
$S_{D}$ & Number of words in the DCM \\
$w_{D}$  & Number of bits in a word in DCM \\
$w_{D}$ & Number of primary input lines \\ 
 $S_{I}$ & Number of words in the IM \\
$w_{I}$ & Number of bits in a word in IM \\\hline
\end{tabular}
 \vspace{-0.8cm}
\end{table}

 The ReVAMP architecture has a three-stage pipeline with instruction fetch~(IF),
 instruction decode~(ID) and execute~(EX) stages.  In the IF stage, the instruction at the address held by the program counter~(PC) is fetched from the IM and loaded into the instruction register~(IR) before  the PC is updated. In the ID stage, the instruction is read from IR to determine the control inputs for the source select multiplexer,  the crossbar interconnect and the write circuit.
  
  The data memory register~(DMR) stores the data read out from the DCM. The primary input register~(PIR) buffers the primary input data. 
  Both DMR and PIR are $w_D$ bits wide. Depending on the control input $M_c$, the source select multiplexer selects either the DMR or the PIR as the data source. Thereafter, 
  the crossbar-interconnect is used to generate the wordline and $w_D$ number of bitline inputs by appropriate permutation of the input data, as per the control signals stored in $C_c$. The 
  crossbar-interconnect is basically a set of multiplixers, one per output, which selects one of the input $w_D$ bits.  The write circuits reads the value of  the target wordline from the register $W_c$ and the output of the crossbar-interconnect to determine and apply the inputs to the row and column decoder of the DCM.


\noindent \textbf{ReVAMP Instruction Set}: The ReVAMP architecture supports two instructions---\textit{Read} and \textit{Apply}, in the formats shown in Fig.~\ref{fig:for}. 
The \textit{Read} instruction reads the word  at the address ${wl}$ from the DCM and stores it in the DMR.
Now available in the DMR, this word can be used as input by the following instructions. 

The \textit{Apply} instruction is used for computation in the DCM. The address ${w}$ specifies the word in the DCM that will be computed upon.
A bit flag ${s}$  chooses whether the inputs will be from primary input~(PIR) or DMR. A two-bit flag ${ws}$ specifies the worline input --- 00 selects logic 0, 01 selects logic 1, 
11 selects input specified by the ${wb}$ flag and 01 is not a valid input.
The ${wb}$ bit-vector are used to specify the bit within the chosen data source for use as wordline input.
Pairs\mbox{~${(v~val)}$}~pairs are used to specify bitline inputs. The bit flag~${v}$ indicates if the input is NOP or a valid input. Similar to ~${wb}$, the  bit-vector~${val}$ 
specifies the bit within the chosen data source for use as bitline input.

\begin{figure}[ht]
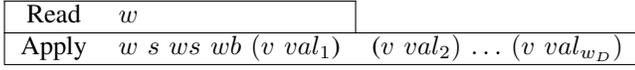

 \begin{tabular}{clcc} \\ \cline{1-2}
\multicolumn{1}{|c}{Read} & \multicolumn{1}{l|}{$w$} &  \\  \cline{1-3}
\multicolumn{1}{|c}{Apply} &$w~s~ws~wb~(v~val_1$) & \multicolumn{1}{c|}{($v~val_2)$ $\ldots$ $(v~val_{w_D})$}   \\ \cline{1-3}
\end{tabular}
\caption{\em ReVAMP instruction format.}
\label{fig:for}
 \vspace{-0.8cm}
\end{figure}
\noindent In each instruction, one bit is required to specify the opcode, and $log_2(S_D)$ bits are required to select the word. 
One bit is required for $s$~flag and two bits are required for the wordline source select flag $ws$. 
Each $($v val$)$ pair requires one bit for the $v$~flag and $log_2(w_D)$ bits for specifying the bit in the selected input source. The  field $wb$ also requires $log_2(w_D)$ bits. Thus, the lengths of these instructions are:
\begin{align}
IL_{Read} &:= 1 + log_2(S_D) \\
 IL_{Apply} &:= 3 + log_2(S_D) +  (1+w_D)(1+ log_2(w_D)) 
\end{align}
The word length $w_I$ of the IM should be greater than or equal to $max(IL_{Read},IL_{Apply})$. 

We demonstate the working of the ReVAMP architecture. Let us consider a $3\times2$ crossbar as the DCM for realizing two-bit XOR function for operands $p_1p_0$ and $q_1q_0$. To compute the XOR, we use the following equation :-
 \begin{align}
p_i \oplus q_i &= p_i.\overline{q_i} + \overline{p_i}.q_i = p_i.\overline{q_i} + \overline{p_i+ \overline{q_i}}
\end{align}
Fig.~\ref{fig:2inp} shows the sequence of operations performed to realize a 2-bit XOR function and the steps are described below. 
\begin{itemize}
	\step{1} Inputs $p_0$ and $p_1$ are loaded to wordline 0 in inverted form via the PIR, since $M_3(0, 1, \overline{p_i}) = \overline{p_i}$.
	\step{2} Wordline 0 is read out using Read instruction. The read out values $\overline{p_0}$ and $\overline{p_1}$ are stored in the DMR. 
	\step{3-4} The read out value is loaded to wordline 1 and 2 using two Apply instructions via the bitlines as $M_3(0, 1, \overline{\overline{p_i}}) = p_i$.
	\step{5} Input $q_0$ and $q_1$ are ANDed  with the values in wordline 2 in inverted form by using 0 as wordline input since $M_3(p_i, 0, \overline{q_i}) = p_i.\overline{q_i}$.
	\step{6} Input $q_0$ and $q_1$ are ORed  with the values in wordline 1 in inverted form by using 1 as wordline input since $M_3(p_i, 1, \overline{q_i}) = p_i+\overline{q_i}$.
	\step{7} The ORed values available in wordline 1 are read out, using Read instructions. 
	\step{8} The values in the DMR are ORed with the contents of wordline 2 to complete the XOR operations, as $p_i.\overline{q_i} + \overline{p_i+ \overline{q_i}}$. 
\end{itemize}

\begin{figure}
    \centering

    \captionsetup[subfigure]{position=b}
    \setlength{\tabcolsep}{1pt}
     \setlength{\extrarowheight}{3pt}
    \begin{subfigure}[t]{\textwidth}

    \caption{ ~~}
    \label{fig:2inp}
\scalebox{0.85}{
\begin{tabular}{cccc cccc cccc ccc}
	& Step 1 &  & ~ &  & Step 3 &  &~  &  & Step 4 &  &~&  & Step 5 &  \\ \cline{2-3} \cline{6-7} \cline{10-11} \cline{13-14} 
	& \multicolumn{1}{|c|}{0} & \multicolumn{1}{|c|}{0} &  & `1' & \multicolumn{1}{|c|}{0} & \multicolumn{1}{|c|}{0} &  &  & \multicolumn{1}{|c|}{$p_1$} & \multicolumn{1}{|c|}{$p_0$} & `0' & \multicolumn{1}{|c|}{$p_1$} & \multicolumn{1}{|c|}{$p_0$} \\  \cline{2-3} \cline{6-7} \cline{10-11} \cline{13-14} 
	& \multicolumn{1}{|c|}{0} & \multicolumn{1}{|c|}{0} &  &  & \multicolumn{1}{|c|}{0} & \multicolumn{1}{|c|}{0} &  & `1' & \multicolumn{1}{|c|}{0} & \multicolumn{1}{|c|}{0} &  & \multicolumn{1}{|c|}{$p_1$} & \multicolumn{1}{|c|}{$p_0$} \\ \hhline{~--~~--}  \cline{10-11} \cline{13-14} 
	`1' & \multicolumn{1}{|c|}{0} & \multicolumn{1}{|c|}{0} &  &  & \multicolumn{1}{|c|}{\ccol $\overline{p_1}$} & \multicolumn{1}{|c|}{\ccol $\overline{p_0}$} &  &  & \multicolumn{1}{|c|}{$\overline{p_1}$} & \multicolumn{1}{|c|}{$\overline{p_0}$} &  & \multicolumn{1}{|c|}{$\overline{p_1}$} & \multicolumn{1}{|c|}{$\overline{p_0}$} \\ \hhline{~--~~--} \cline{10-11} \cline{13-14} 
	& $p_1$ & $p_0$ &  &  & $\overline{p_1}$ & $\overline{p_0}$ &  &  & $\overline{p_1}$ & $\overline{p_0}$ &  &  $q_1$& $q_0$ \\ 
	&  &  &  &  &  &  &  &  &  &  &  &  &  \\
	& Step 6 &  &  &  & Step 8 &  &  &  &  &  &  &  &  \\ \cline{2-3} \cline{6-7} \cline{10-11} 
	& \multicolumn{1}{|c|}{$p_1.\overline{q_1}$} & \multicolumn{1}{|c|}{$p_0.\overline{q_0}$} &  & `1' & \multicolumn{1}{|c|}{$p_1.\overline{q_1}$} & \multicolumn{1}{|c|}{$p_0.\overline{q_0}$} &  &  & \multicolumn{1}{|c|}{$p_1\oplus q_1$} & \multicolumn{1}{|c|}{$p_0 \oplus q_0 $} &  &  &  \\\hhline{~--~~--} \cline{10-11} 
	`1' & \multicolumn{1}{|c|}{$p_1$} & \multicolumn{1}{|c|}{$p_0$} &  &  & \multicolumn{1}{|c|}{\ccol $p_1+\overline{q_1}$} & \multicolumn{1}{|c|}{\ccol $p_0+\overline{q_0}$} &  &  & \multicolumn{1}{|c|}{$p_1+\overline{q_1}$} & \multicolumn{1}{|c|}{$p_0+\overline{q_0}$} &  &  &  \\ \hhline{~--~~--}  \cline{10-11} 
	& \multicolumn{1}{|c|}{$\overline{p_1}$} & \multicolumn{1}{|c|}{$\overline{p_0}$} &  &  & \multicolumn{1}{|c|}{$\overline{p_1}$} & \multicolumn{1}{|c|}{$\overline{p_0}$} &  &  & \multicolumn{1}{|c|}{$\overline{p_1}$} & \multicolumn{1}{|c|}{$\overline{p_0}$} &  &  &  \\ \cline{2-3} \cline{6-7} \cline{10-11} 
	& $q_1$ & $q_0$ &  &  & $p_1+\overline{q_1}$ & $p_0+\overline{q_0}$ &  &  &  &  &  &  &  \\
\end{tabular}}
 \end{subfigure}
\\ %
\begin{subfigure}[t]{0.5\textwidth}
\centering
\caption{ ~~}
\label{fig:2inp_ins}
\footnotesize
\begin{tabular}{rl}\bottomrule
 & \textbf{Instruction} \\ \midrule
$I_1$ & Apply 0 0 01 1 0 1 1 \\
$I_2$ & Read 0 \\
$I_3$ & Apply 2 1 01 1 0 1 1 \\
$I_4$ & Apply 1 1 01 1 0 1 1 \\ 
$I_5$ & Apply 2 0 00 1 0 1 1 \\
$I_6$ & Apply 1 0 01 1 0 1 1 \\
$I_7$ & Read 1 \\
$I_8$ & Apply 2 1 01 1 0 1 1 \\ \toprule
\end{tabular}
\end{subfigure}

    \caption{\em XOR computation of two-bit vectors $p_1p_0$ and $q_1q_0$. (a) The steps of computation are shown graphically using a crossbar schema. The read out steps are not shown explicitly. The word highlighted in green represents the read out word. (b) The corresponding instructions for the ReVAMP architecture. The inputs
    $p_1p_0$ and $q_1q_0$ are made available on the PIR during Step 1 and Step  5-6 respectively.}
    \label{fig:2inpxor}
     \vspace{-0.8cm}
\end{figure}

\noindent The set of instructions corresponding to the steps is shown in~Fig.~\ref{fig:2inp_ins}. This concludes the description of the ReVAMP architecture. In the following subsection, we describe briefly structural representation of Boolean functions.

\subsection{Logic representation}
\noindent For representation of Boolean functions, we use two structural representations namely And Inverter Graph~(AIG)~\cite{mishchenko2006using} and Majority Inverter Graph~(MIG)~\cite{amaru2016majority}. 
An AIG~(MIG) is a directed acyclic graph where each node is 2-input~(3-input) representing Boolean AND~(Boolean Majority). 
A directed edge \mbox{$i\rightarrow j$} exists if the output of the (\textit{parent}) node $i$ is an input to the (\textit{child}) node $j$. Each edge is marked as either regular or inverted. A Primary Input  (PI) node is either a logic constant $0/1$ or a Boolean variable. If a node is not a PI, then it is an \textit{internal} node. A Primary Output~(PO) node represents the output of the function. An AIG~(MIG) can have one or more PO nodes. We define the level of a node $n$ as follows.
\begin{definition}
	The level of a node $n$, written as  $level(n)$, is defined as the length of the longest path from any PI node to the node $n$. The level of  the PI nodes is zero. 
\end{definition}

\begin{example}
 Fig.~\ref{fig:aig} and Fig.~\ref{fig:mig} shows an AIG and a MIG respectively. In both the graphs, the primary inputs~($a_0, a_1, a,b, \ldots$) are shown in square boxes and the internal nodes~($n1,n2,S_1,S_2, \ldots$)  are shown in circles. The output nodes~($n5$,$S_4$) are shown in double lined circles. The inverted edges~($n2 \rightarrow n4, S_2 \rightarrow S_3, \ldots$) are marked using dots.
\end{example}

\begin{figure}[ht]

\begin{subfigure}[t]{0.5\textwidth}
\caption{~}
\label{fig:aig}
\centering
\includegraphics[width=2.2in]{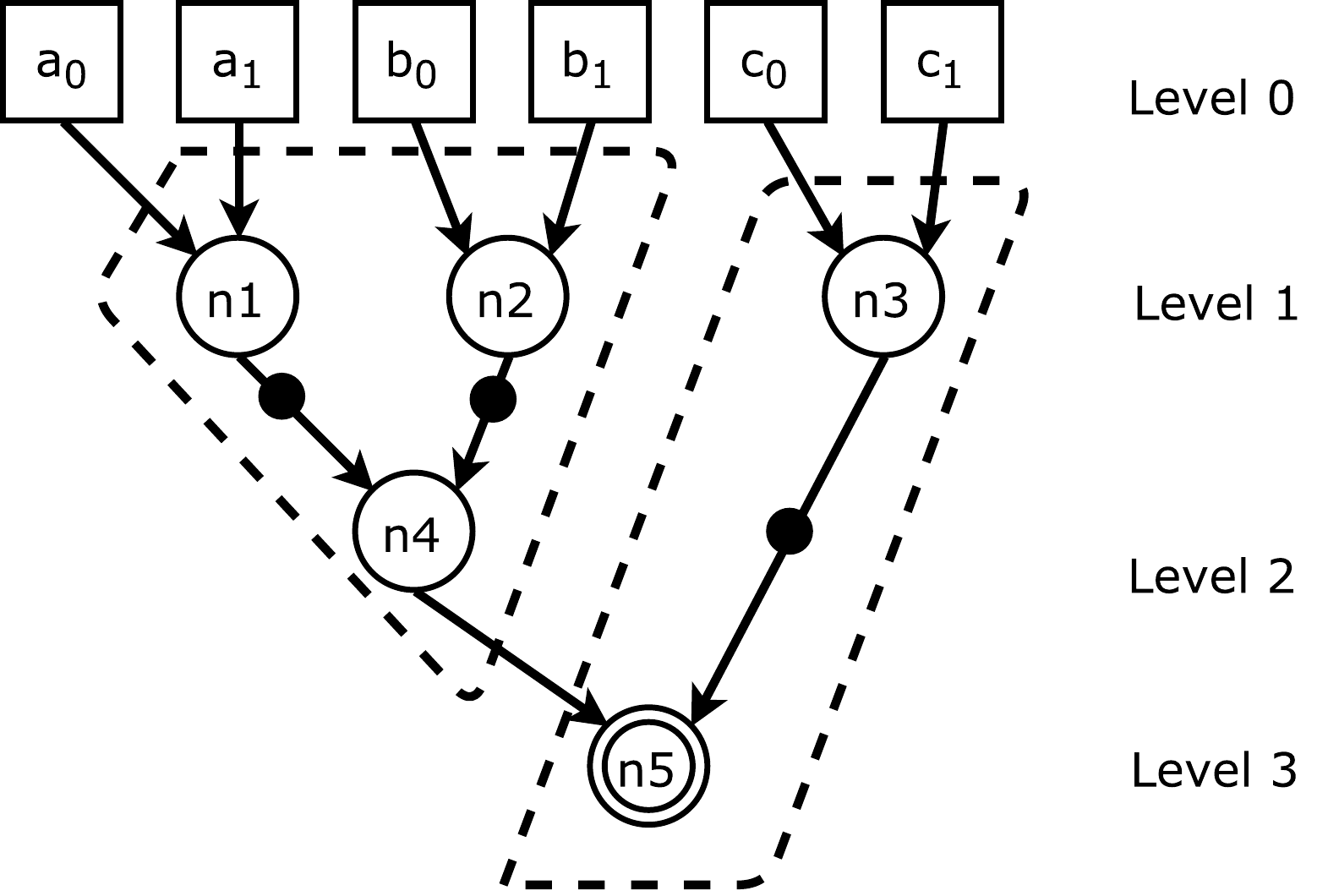}
 \end{subfigure}
\\ %
\begin{subfigure}[t]{0.5\textwidth}
\centering
\caption{~}
\label{fig:mig}
\includegraphics[width=2.2in]{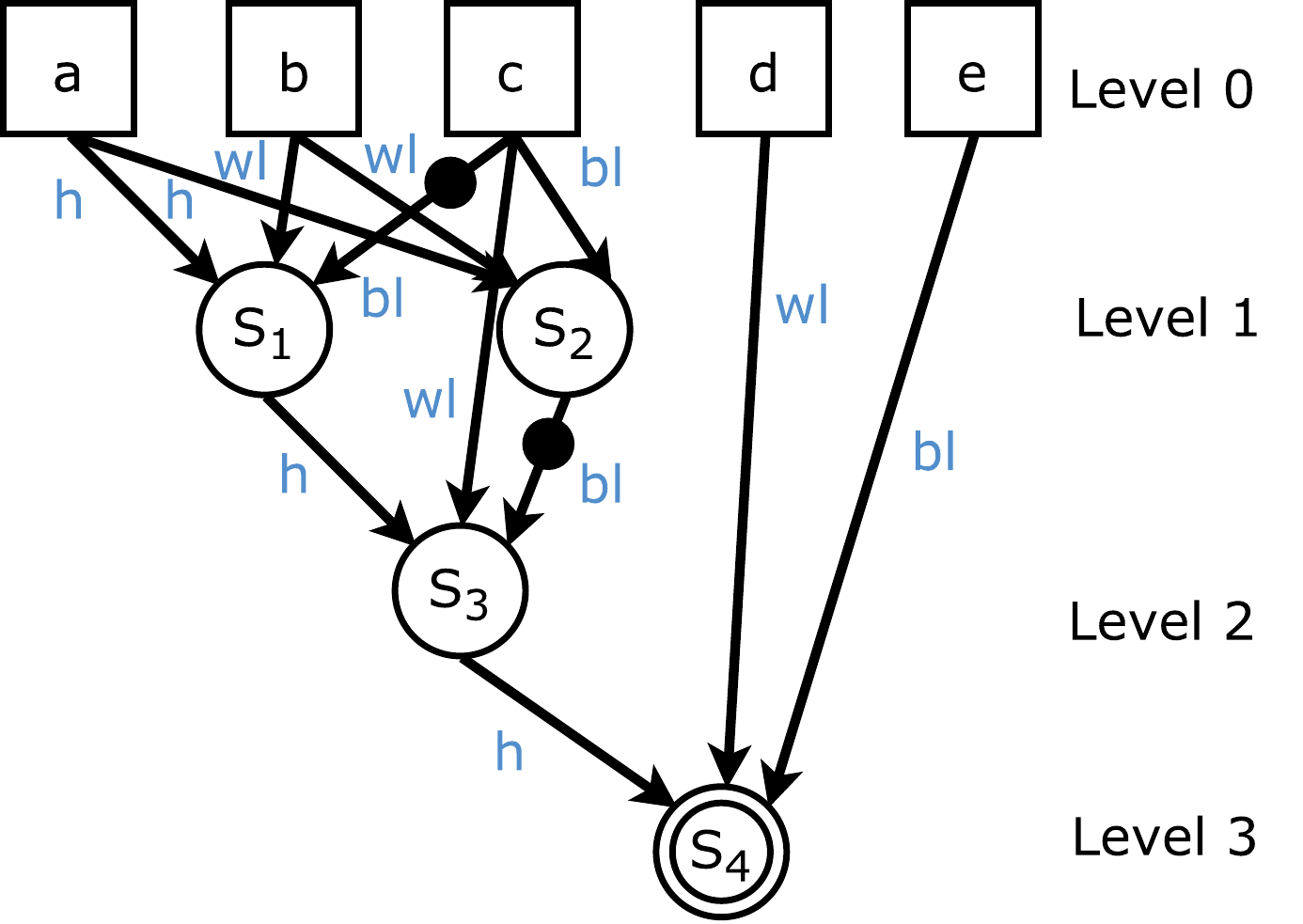}
\end{subfigure}
\caption{\em Logic representation. (a) An AIG (b) A MIG.}
\label{fig:graph}
 \vspace{-0.8cm}
\end{figure}

%
%
%
%
%
%
%
%
%
%

  \section{Problem definition and Solution}\label{sec:problem}
\noindent In this section, we present the technology mapping problem for the ReVAMP architecture along with overview of the proposed solutions.
\subsection{Problem definition}
\noindent {\em Area constrained techhnology mapping :} Given a Boolean function represented as a Boolean logic network $G$ and crossbar dimension $S_D\times w_D$, determine a sequence of instructions $I_1, I_2, ..., I_T$, $I_t \in \{Read, Apply\}$ and $1 \le t \le T$
and PIR inputs for the ReVAMP architecture that computes the 
output nodes of the network $G$. \\
\noindent {\em Delay focused technology mapping :} Given a Boolean function represented as a Boolean logic network $G$ and crossbar width $w_D$, determine a sequence of instructions $I_1, I_2, ..., I_T$, $I_t \in \{Read, Apply\}$ and $1 \le t \le T$, PIR inputs and number of words $S_D$ for the ReVAMP architecture that computes the output nodes of the network $G$. 

\noindent The quality of the solution is measured in terms of the {\em delay} and the {\em total number of devices} required for the mapping. 
The {\em delay} of a solution is equal to the number of instructions~($T$). The total number of devices is equal to $S_D \times w_D$.

\subsection{Solution approach}
\noindent In this paper, we propose two different approaches to the problem. Fig.~\ref{fig:flowchart} shows the overall flowchart for the technology mapping problem. 
In the first approach, we consider the area constrained version of the problem, where we represent the Boolean function as an AIG. We begin by partitioning the AIG into $k$-input Look-up Tables~(LUTs). A $k$-input LUT is basically a function with atmost $k$-inputs and a single output. Once the graph has been partitioned, the LUTs for computation are scheduled in topological ordering, i.e. the LUTs 
close to the primary input are scheduled first and so on, till the output LUTs are computed. In order to compute a LUT, we express the functionality of the LUT using Exclusive Sum-Of-Products~(ESOP)~\cite{mishchenko2001fast}. 
Any arbitrary ESOP can be computed on the DCM with at least 3-wordlines and 2-bitlines~(explained in detail in Theorem~\ref{thm:32}) --- the variables which have to be used in inverted form are negated first~(to be applied via bitlines), followed by computing the product terms and finally XORing them. To reduce the delay, the AND computation for realizing the ESOP needs to minimize number of reads performed and maximize number of AND operations that can be done in parallel. Thereafter, we perform the XOR of computed AND terms by means of a XOR reduction tree of logarithmic depth in the number of AND terms.

In the second approach, we focus on minimizing the {\em delay} of the mapping, without any constraints on the number of words.  We use MIGs for logic representation in this approach.  
We propose an algorithm  with four phases --- assignment of nodes as host or input for computation,  grouping nodes to blocks, packing blocks to
words followed by generation and scheduling of instructions. We explain both the technology mapping solutions in detail in the following sections.
\begin{figure}[t]
\centering
\includegraphics[width=0.4\textwidth]{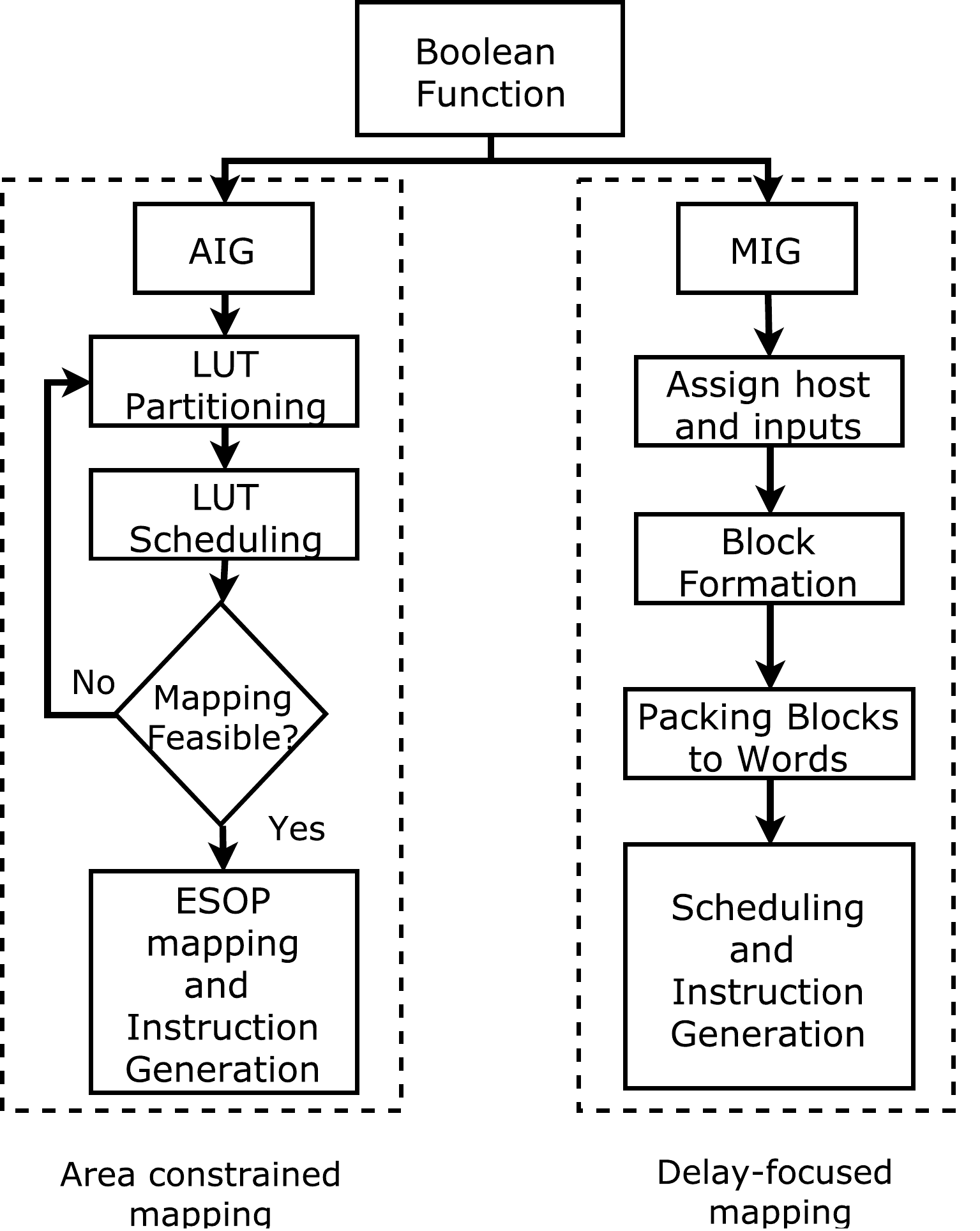}
\vspace{0.3cm}
\caption{\em Technology mapping flow for the ReVAMP architecture}
\label{fig:flowchart}
 \vspace{-0.8cm}
\end{figure}

  \section{Area-constrained technology mapping}\label{sec:area}
\noindent In this section, we establish a lower bound on the
number of devices required to map any arbitrary AIG~(MIG). Thereafter, we present a scalable technique for area constrained technology
mapping.

\begin{theorem}
Any AIG or MIG with $k$-levels can be mapped using $2(k+1)$ devices, arranged as a crossbar with atleast two bitlines.
\end{theorem}
\noindent \textbf{Proof:} Since any AIG can be expressed as MIG, we prove the theorem for
MIG by means of an inductive proof. Before explaining the proof,
we describe a transformation to the input MIG and prove the theorem on the transformed MIG. We transform the MIG such that
\begin{itemize}
\item Each internal node has a single child. Nodes with multiple fanout can be replicated bottom-up i.e., from the output to the primary inputs. 
\item Each node has two non-inverted inputs and a single inverted input.  This can be realized by propagating the inverts across nodes or by creating an inverted copy of the node as required, using the following axiom for Boolean
majority.
\begin{align}
 \overline{M_3(a,b,c)} = M_3(\overline{a},\overline{b},\overline{c})
\end{align}
\end{itemize}
Now, we present the inductive proof for the transformed MIG. The device at wordline 0 bitline 0 is used for inverting any input $v$ as needed by applying the input via the bitline with 1 as wordline input and 0 as internal state, i.e. $M_3(0,1,\overline{v})$.  
The inverted value $\overline{v}$ can be read out in the next cycle and used in the following cycles using Apply instructions. Any device is reset i.e. internal state $Z$ is set to logic 0, by applying 0 and 1 as wordline and bitline input
respectively. \\

\noindent \textbf{Base Case:} A MIG with 1-level basically implies inputs act as outputs and hence does not require any devices for computation. Therefore, we consider the MIG in Fig.~\ref{fig:kdevicea} with 2-levels as the base case. One of the non-inverted input $W$ and the inverted input  $B$ can be loaded to wordline 1. The second non-inverted input $H$ is loaded to wordline 0. The wordline 1 can be readout and in the next cycle, $W$ and $B$ are applied as wordline and bitline inputs of the device holding $H$ to compute $S$.

\noindent \textbf{Inductive Case:} Let us assume that for an MIG with $k$-levels, the theorem holds true. Now, consider an MIG with $k+1$-levels, as shown in Fig.~\ref{fig:kdeviceb}. The subtrees $MIG_{wk}$, $MIG_{bk}$ and $MIG_{hk}$ have $k$-levels. Therefore, these MIGs can be computed using $2(k+1)$ devices. Let the subtree $MIG_{wk}$  be computed on wordlines $1$ to $(k+1)$ and the result $W_k$ be stored at wordline $1$ bitline $1$. All the devices, except the device holding $W_k$ is reset. Similarly, subtree $MIG_{bk}$  be computed on wordlines $1$ to $k+1$ and the result $B_k$ is stored at wordline $1$ bitline $0$, followed by reset of all the devices, except wordline 1. The last subtree
$MIG_{wk}$ is computed using wordlines $0$, $2$ to $k+1$ with the result $H_k$ stored at wordline $1$ and bitline $1$. Therefore, to compute the final output $T_{k+1}$, wordline 1 is read out 
and then $W_k$ and $B_k$ are applied to the wordline and bitline of device holding $H_k$ to compute $T_{k+1}$.  This completes the proof. $\blacksquare$

\begin{figure}[t]
\centering
\begin{subfigure}[t]{3.8in}
	\caption{~~}
	\label{fig:kdevicea}
	\includegraphics[width=3in]{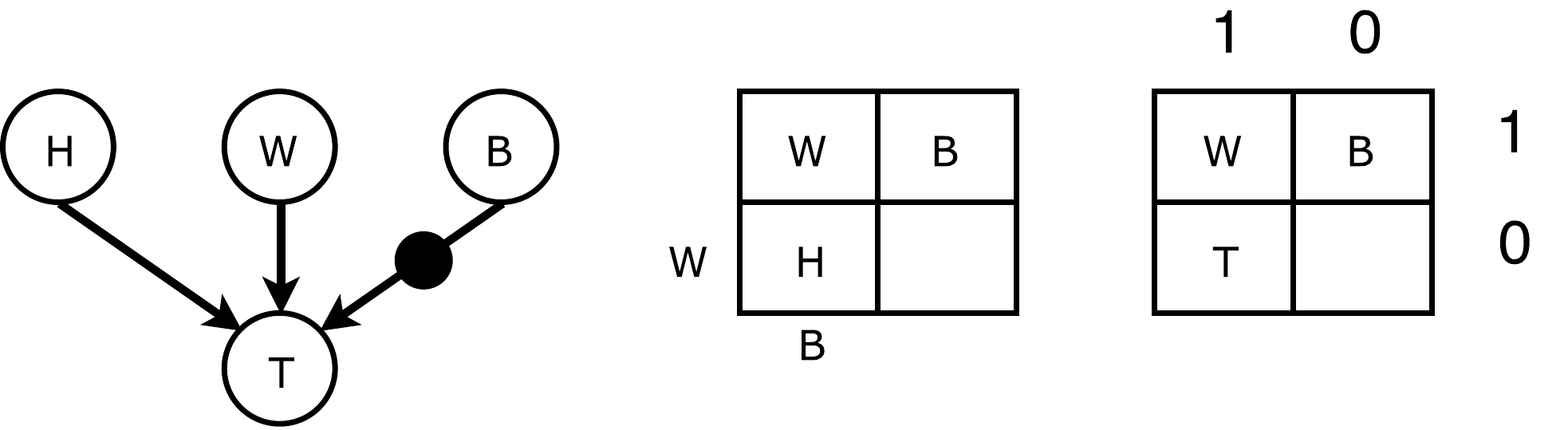}
	\end{subfigure}\\
	\begin{subfigure}[t]{3.8in}
	\caption{ ~~}
	\label{fig:kdeviceb}
	\includegraphics[width=3in]{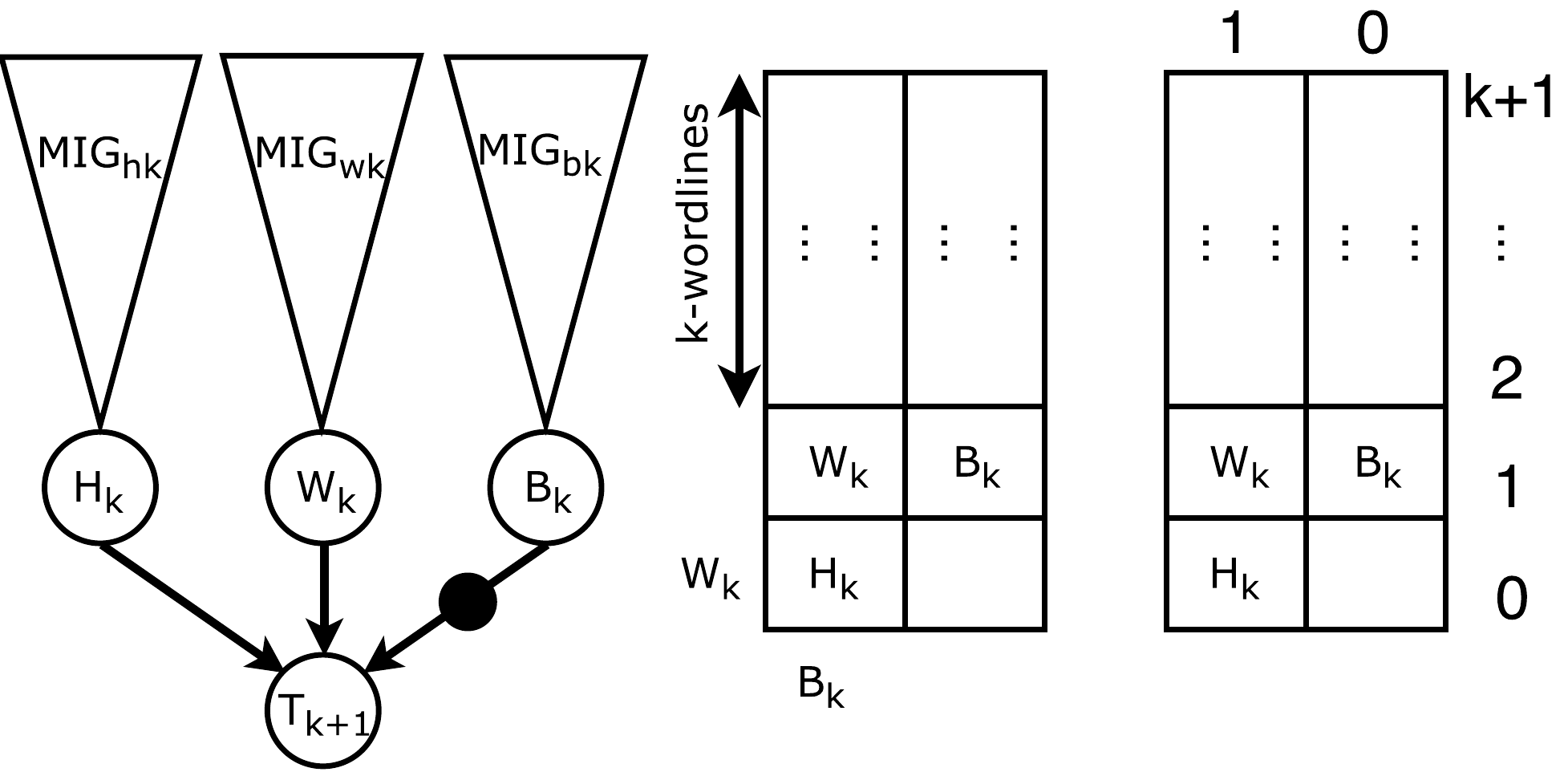}
	\end{subfigure}
	\caption{\em  (a) Mapping an MIG with 2-levels  (b) Mapping an MIG with $(k+1)$-levels.}
\label{fig:kdevice}
\end{figure}
\subsection{Logic network partitioning and scheduling}
\noindent We represent the Boolean function as an AIG. We partition the graph into $k$-input LUTs using ABC~\cite{abcalan}. From here on, we refer to the partitioned graph as LUT graph and each node in the partitioned graph represents a LUT. 
\begin{example}
 For $k=4$, the AIG in Fig.~\ref{fig:aig} can be partitioned into two LUTs, as shown by dotted lines. 
\end{example}
\noindent{\em Bound on number of devices required :} To determine the bound on number of devices required for the storage of intermediate results, we define {\em transient node}.
\begin{definition}
In a LUT graph, a node $n$ is termed as transient node in level $l$ if node $level(n) < l$  and there exists an edge, $n \rightarrow n'$  such that $level(n') > l$.   
\end{definition}

\begin{figure}[t]
\centering
\includegraphics[width=0.4\textwidth]{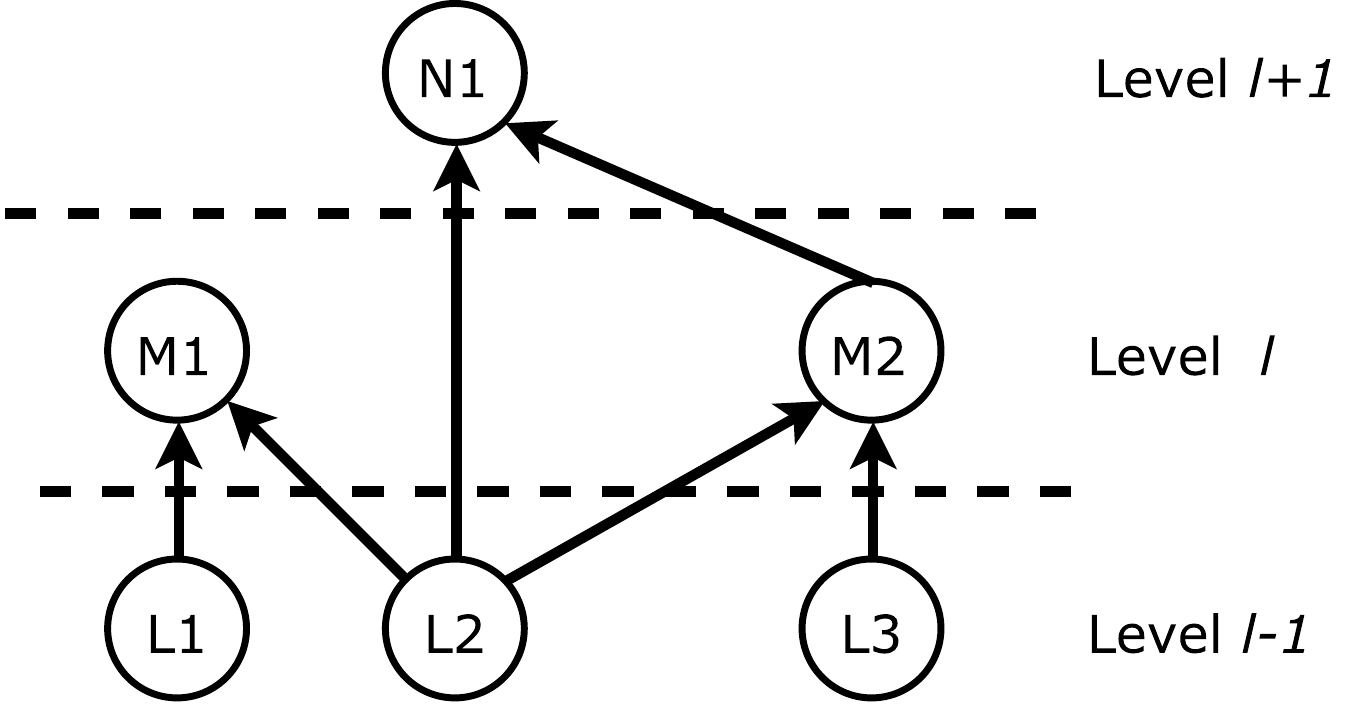}
\caption{\em A portion of the LUT graph. Each node in this partitioned graph represents a LUT.}
\label{fig:transient}
 \vspace{-0.8cm}
\end{figure}

\begin{figure}[t]
\centering
\includegraphics[width=2in]{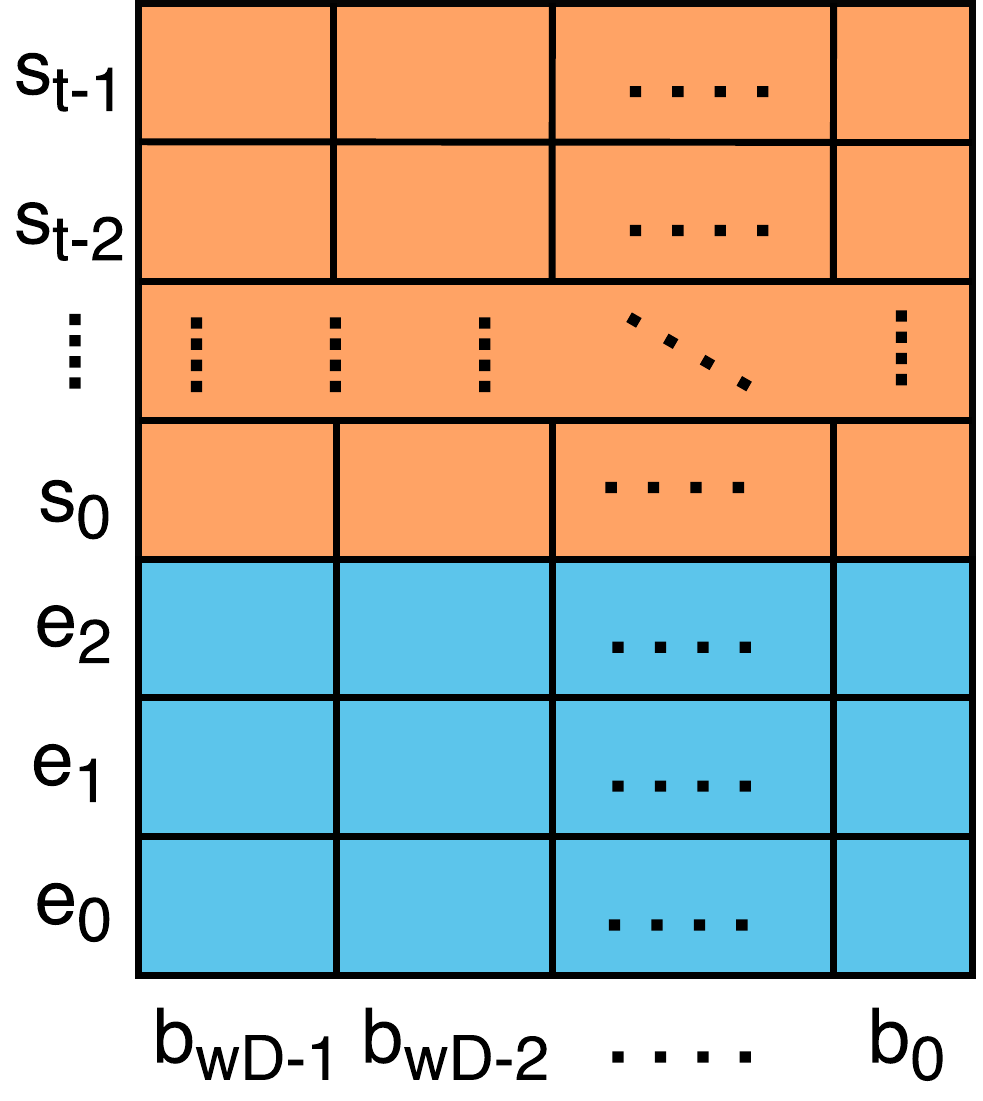}
\caption{\em Memory layout of the ReRAM crossbar array with $r$ wordlines and $c$ bitlines. The three wordlines $e_0$, $e_1$ and $e_2$ are used for computation. The rest of the wordlines  $s_0, \ldots, s_t$ are used for storage of the intermediate results. $3+t = S_D$. }
\label{fig:mem}
 \vspace{-0.8cm}
\end{figure}
\begin{example}
In Fig.~\ref{fig:transient}, LUT~$L2$ in level~$l-1$ has an edge to LUT~$N1$ in level~$l+1$, therefore it is a transient node for level~$l$.
\end{example}
Let the number of nodes, including transient nodes in a level $l$ be $N_l$. 
We can schedule the nodes of the LUT Graph in topological ordering, i.e all nodes at level~$l-1$ are scheduled before any node in level~$l$ is scheduled. A node in level~$l$ is dependent only on the nodes~(including transient nodes) that are present in level~$l-1$. Therefore, once all the nodes in level~$l$ have been scheduled, the nodes in level~$l-1$ can be reset. Doing this iteratively, the number of devices required for scheduling a LUT graph is 
\begin{align}\label{eq:mindev}
Min_{Dev} &= \underset{0 \le l \le L_{max}-1}{max}(N_l+N_{l+1}) 
\end{align} 
\noindent The memory layout of the crossbar, with \mbox{$S_D$~($=t+3$) wordlines} and $w_D$ bitlines is shown in Fig.~\ref{fig:mem}. The top $t$ wordlines are used for storing the output of each LUT. 
The bottom three wordlines $e_0, e_1$ and $e_2$ are reserved for computation of each LUT. For the  scheduling to be feasible,  $Min_{Dev}$  should be less than or equal to $ (t\times w_D)$.
If the scheduling condition is not feasible, a different value of $k$ is used to partition the graph and feasibility is checked. Once the scheduling condition is satisfied for a given crossbar size, nodes are scheduled in topological order. The device where the output of an LUT~(node in the LUT graph) would be stored,  is determined according to the best fit method. 
The wordline in the crossbar with minimum number of free devices is chosen if the number of nodes to  schedule is less than or equal to the number of free devices in that wordline.
If no such wordline exists, a wordline with maximum number of free devices is chosen iteratively, till all the nodes have been allocated a device. A device storing a node $n$
is marked {\em dirty} if all the successors of $n$ have already been allocated. If none of the devices are {\em free}, then the wordline with maximum dirty bits is reset and allocation starts. This process is repeated till all the nodes have been scheduled, along with target device allocation.  The overall technique has been shown in Algorithm~\ref{algo:scheduling}.

\begin{algorithm}
\scriptsize
 \KwData{Lut Graph $G$,$S_D$,$w_D$}

 \For{$l=1; l \le l_{max}; l$++}
 {

        Allocate unscheduled nodes in level $l$ to free devices in a wordline considering Best-fit\;
        \If{$s.scheduled = True$ : $\forall s \in succ(n)$}
        {
	  \tcp{Device allocated to node $n$ is marked dirty}
	  $dev(n).dirty = True$\;
        }
     \While{$level(n) = l$ and $n.scheduled = False$ : $\exists n \in G(V)$}
     {
	 \tcp{ No free device is available}
	 \If{$\nexists free(D) $}
	{
	 $w$ = wordline with maximum number of {\it dirty} devices\;
	 Reset the {\it dirty} devices in wordline $w$\;
	}
	Allocated unscheduled nodes of level $l$ to the free devices in wordline~$w$\;
     }
 }
 
 \caption{LUT Graph Scheduling}
 \label{algo:scheduling}
\end{algorithm}

\begin{figure}[ht]
	\begin{center}
		\includegraphics[width=2in]{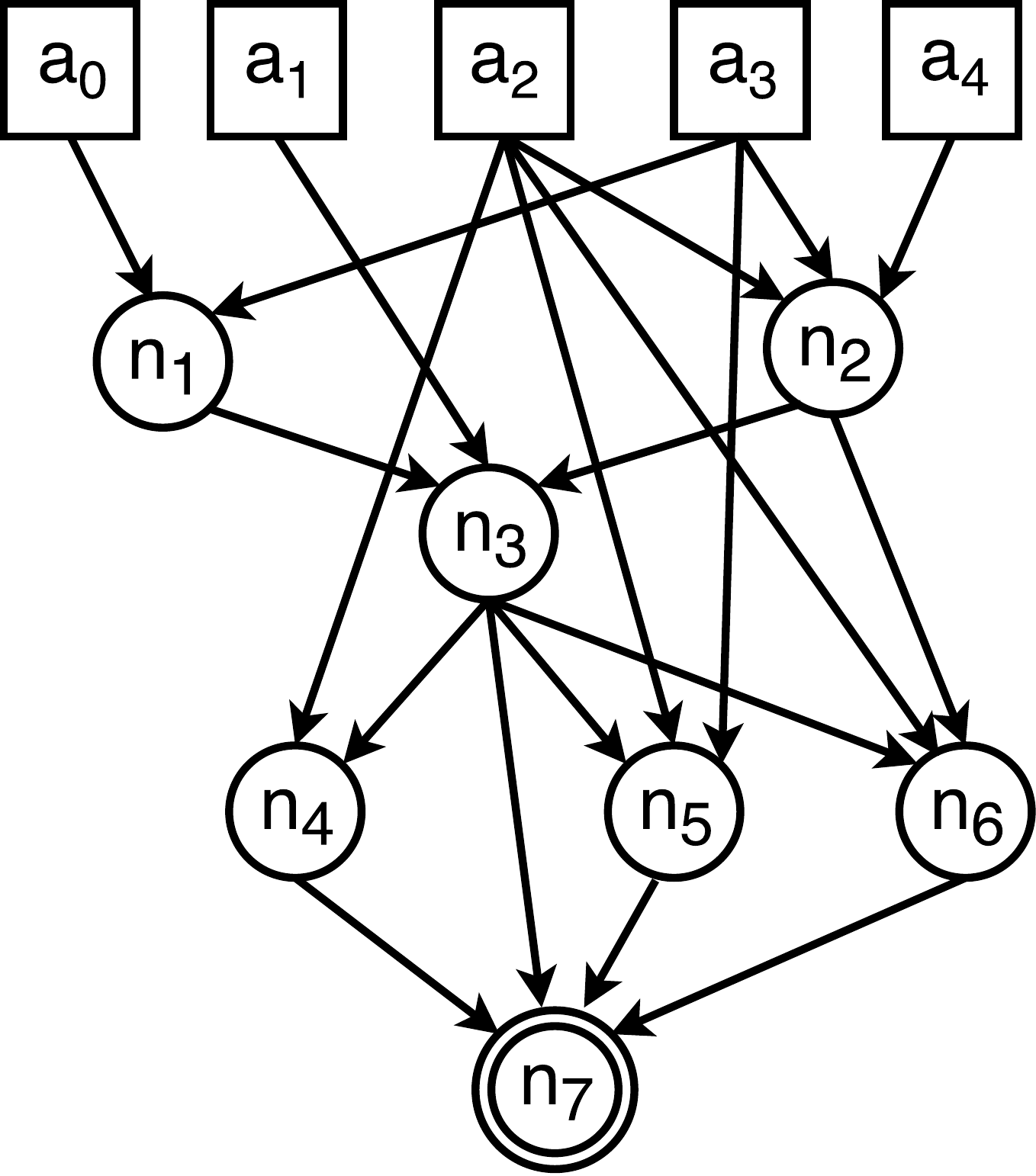}
	\end{center}
	\caption{\em An LUT graph with five primary inputs~($a_0,\dots,a_5$) with LUT nodes~($n_1, \ldots, n_7$). The output of LUT $n_7$ is the output of the LUT graph. }
	\label{fig:lutsch}
	 \vspace{-0.6cm}
\end{figure}

\begin{figure}[ht]
	
	\centering
	\captionsetup[subfigure]{position=b}
	\setlength{\tabcolsep}{4pt}
	\setlength{\extrarowheight}{3pt}
	{\scriptsize
		\scalebox{0.9}{
			\begin{tabular}{cccc cccc cccc cccc}
				& \multicolumn{4}{c}{Inital State}  &  &  \multicolumn{4}{c}{Level 1}  &  &  \multicolumn{4}{c}{Level 2}   & \\ \hhline{~----} \cline{7-10} \cline{12-15} 
				& \multicolumn{1}{|c|}{\comp $0$} & \multicolumn{1}{|c|}{\comp$0$}  & \multicolumn{1}{|c|}{\comp$0$}  & \multicolumn{1}{|c|}{\comp$0$}
				& ~~ & \multicolumn{1}{|c|}{$0$} & \multicolumn{1}{|c|}{$0$} & \multicolumn{1}{|c|}{$\overline{n_2}$}  & \multicolumn{1}{|c|}{$\overline{n_1}$}
				& ~~ &  \multicolumn{1}{|c|}{$0$} & \multicolumn{1}{|c|}{$\overline{n_3}$}  & \multicolumn{1}{|c|}{$\overline{n_2}$}  & \multicolumn{1}{|c|}{$\overline{n_1}$}
				\\ \hhline{~----} \cline{7-10} \cline{12-15}
				& \multicolumn{1}{|c|}{\comp$0$} & \multicolumn{1}{|c|}{\comp$0$}  & \multicolumn{1}{|c|}{\comp$0$}  & \multicolumn{1}{|c|}{\comp$0$}
				& &\multicolumn{1}{|c|}{$0$} & \multicolumn{1}{|c|}{$0$}  & \multicolumn{1}{|c|}{$0$}  & \multicolumn{1}{|c|}{$0$}
				& & \multicolumn{1}{|c|}{$0$} & \multicolumn{1}{|c|}{$0$}  & \multicolumn{1}{|c|}{$0$}  & \multicolumn{1}{|c|}{$0$}
				\\ \hhline{~----}  \cline{7-10} \cline{12-15}
				& \multicolumn{1}{|c|}{\res$0$} & \multicolumn{1}{|c|}{\res$0$}  & \multicolumn{1}{|c|}{\res$0$}  & \multicolumn{1}{|c|}{\res$0$}
				& &\multicolumn{1}{|c|}{$0$} & \multicolumn{1}{|c|}{$0$}  & \multicolumn{1}{|c|}{$0$}  & \multicolumn{1}{|c|}{$0$}
				& & \multicolumn{1}{|c|}{$0$} & \multicolumn{1}{|c|}{$0$}  & \multicolumn{1}{|c|}{$0$}  & \multicolumn{1}{|c|}{$0$}
				\\ \hhline{~----}  \cline{7-10} \cline{12-15}
				& \multicolumn{1}{|c|}{\res$0$} & \multicolumn{1}{|c|}{\res$0$}  & \multicolumn{1}{|c|}{\res$0$}  & \multicolumn{1}{|c|}{\res$0$}
				& &\multicolumn{1}{|c|}{$0$} & \multicolumn{1}{|c|}{$0$}  & \multicolumn{1}{|c|}{$0$}  & \multicolumn{1}{|c|}{$0$}
				& & \multicolumn{1}{|c|}{$0$} & \multicolumn{1}{|c|}{$0$}  & \multicolumn{1}{|c|}{$0$}  & \multicolumn{1}{|c|}{$0$}
				\\ \hhline{~----}  \cline{7-10} \cline{12-15}
				& \multicolumn{1}{|c|}{\res$0$} & \multicolumn{1}{|c|}{\res$0$}  & \multicolumn{1}{|c|}{\res$0$}  & \multicolumn{1}{|c|}{\res$0$}
				& &\multicolumn{1}{|c|}{$0$} & \multicolumn{1}{|c|}{$0$}  & \multicolumn{1}{|c|}{$0$}  & \multicolumn{1}{|c|}{$0$}
				& & \multicolumn{1}{|c|}{$0$} & \multicolumn{1}{|c|}{$0$}  & \multicolumn{1}{|c|}{$0$}  & \multicolumn{1}{|c|}{$0$}
				\\ \cline{2-5}  \cline{7-10} \cline{12-15}
				
				& \multicolumn{4}{c}{Level 3}  &  &  \multicolumn{4}{c}{Level 4}  &  &  \multicolumn{4}{c}{}   & \\ \cline{2-5} \cline{7-10} 
				&  \multicolumn{1}{|c|}{$0$} & \multicolumn{1}{|c|}{$\overline{n_3}$}  & \multicolumn{1}{|c|}{$\overline{n_2}$}  & \multicolumn{1}{|c|}{$\overline{n_1}$}
				& & \multicolumn{1}{|c|}{$\overline{n_7}$} & \multicolumn{1}{|c|}{$\overline{n_3}$}  & \multicolumn{1}{|c|}{$\overline{n_2}$}  & \multicolumn{1}{|c|}{$ \overline{n_1}$}
				\\ \cline{2-5}  \cline{7-10} 
				& \multicolumn{1}{|c|}{0} & \multicolumn{1}{|c|}{$\overline{n_6}$}  & \multicolumn{1}{|c|}{$\overline{n_5}$}  & \multicolumn{1}{|c|}{$\overline{n_4}$}
				& &\multicolumn{1}{|c|}{0} & \multicolumn{1}{|c|}{$\overline{n_6}$}  & \multicolumn{1}{|c|}{$\overline{n_5}$}  & \multicolumn{1}{|c|}{$\overline{n_4}$}
				\\ \cline{2-5}  \cline{7-10} 
				& \multicolumn{1}{|c|}{$0$} & \multicolumn{1}{|c|}{$0$}  & \multicolumn{1}{|c|}{$0$}  & \multicolumn{1}{|c|}{$0$}
			    & &\multicolumn{1}{|c|}{$0$} & \multicolumn{1}{|c|}{$0$}  & \multicolumn{1}{|c|}{$0$}  & \multicolumn{1}{|c|}{$0$}
				\\ \cline{2-5}  \cline{7-10} 
				& \multicolumn{1}{|c|}{$0$} & \multicolumn{1}{|c|}{$0$}  & \multicolumn{1}{|c|}{$0$}  & \multicolumn{1}{|c|}{$0$}
				& &\multicolumn{1}{|c|}{$0$} & \multicolumn{1}{|c|}{$0$}  & \multicolumn{1}{|c|}{$0$}  & \multicolumn{1}{|c|}{$0$}
				\\ \cline{2-5}  \cline{7-10} 
				& \multicolumn{1}{|c|}{$0$} & \multicolumn{1}{|c|}{$0$}  & \multicolumn{1}{|c|}{$0$}  & \multicolumn{1}{|c|}{$0$}
				& &\multicolumn{1}{|c|}{$0$} & \multicolumn{1}{|c|}{$0$}  & \multicolumn{1}{|c|}{$0$}  & \multicolumn{1}{|c|}{$0$}
				\\ \cline{2-5}  \cline{7-10} 
				
			\end{tabular}
		}
		
		\medskip			
	}

	\caption{\em Scheduling using Algorithm~\ref{algo:scheduling} for the LUT graph in Fig.~\ref{fig:lutsch}. The wordlines 0-3 (colored in teal) are reserved for XOR computation and the remaining wordlines 4-5 (colored in orange) are used for storage of the LUT outputs.}
	\label{fig:3cat_2}
\end{figure}
\begin{example} We explain the device allocation and scheduling technique, presented in Algorithm~\ref{algo:scheduling} using a representative LUT graph, shown in Fig.~\ref{fig:lutsch}. The nodes are scheduled in topological ordering. Nodes in level 1, $n_1$ and $n_2$, are allocated to wordline 5, as shown in Fig.~\ref{fig:3cat_2}. Node $n_3$, in level 2, is assigned another device in wordline 5, using the Best-fit allocation strategy. Since the only successor of $n_1$ has been allocated, device allocated to $n_1$ is now marked dirty. In level 3, there are 3 nodes~($n_4, n_5$ and $n_6$). Since there is only a single device free in wordline 5, it is not possible to allocate these nodes together. Therefore, these nodes are allocated to wordline 4. All the successors of node $n_2$ have been allocated, hence the corresponding device is marked as dirty. Finally, the node $n_7$ in level 4 is allocated to the free device in wordline 5. This completes the allocation and scheduling of the LUT nodes.
\end{example}

\subsection{ESOP computation}
\noindent Each function realized by the LUT can be expressed as an Exclusive Sum-Of-Product~(ESOP). For many Boolean functions, minimal ESOPs have lesser number of cubes compared to Sum-Of-Products~\cite{sasao1996representations}. In addition, there are multiple ESOP minimizers available which can be used to reduce the ESOP size~\cite{kozlowski1995enhanced,drechsler1995sympathy,zakrevskij1995minimum,drechsler1999pseudo}. Before presenting the ESOP computation algorithm on ReVAMP, we
present a brief description of the related terms.
\begin{definition}
 A {\em literal} is a Boolean variable either in inverted or non-inverted form.
\end{definition}
\begin{definition}
 A {\em cube} is a product term composed of 
literals using Boolean AND. 
\end{definition}
\begin{example}
 The ESOP $\overline{a}bc \oplus a\overline{b}c$ has two cubes, $\overline{a}bc$ and $a\overline{b}c$.  The cube $\overline{a}bc$ has literal $a$ in inverted form and $b$,$c$ in non-inverted form.
\end{example}
\noindent 
\begin{theorem}\label{thm:32}
 Any Boolean function, expressed as an ESOP, can be computed using three wordlines and atleast two bitlines. 
\end{theorem}
\noindent \textbf{Proof~:} We present a constructive proof for the theorem. Let us consider three wordlines, $e_0$, $e_1$ and $e_2$ with bitlines $b_0$ and $b_1$. We consider two cases. \\
{\em Case 1:} The ESOP has a single cube, say $l_1.l_2...l_n$. If a literal $l_i$ is inverted, it is applied via bitline $b_0$ with `0' as input to wordline $e_2$. Else, the literal is applied via bitline $b_0$ and `1' as input to wordline $e_0$ to store in non-inverted form. Then, wordline~$e_0$ is read out and $\overline{l_i}$ is applied via the bitline with `0' as wordline input to wordline~$e_2$. 
The wordline $e_0$ is reset. The process is repeated till all the literals have been ANDed and the computed cube is available at wordline~$e_2$ and bitline~$b_0$. \\
{\em Case 2:} The ESOP has more than one cube, say $c_1, c_2,...,c_m$. The cube $c_1$ can be computed, as stated in {\em Case 1}. 
Similarly, $c_2$ can be computed  at wordline $e_2$ and bitline $b_1$ by applying the bitline inputs via bitline $b_1$. The cubes $c_1$ and $c_2$ can be XORed as shown in Fig.~\ref{fig:proofcube} with the result stored at wordline $e_2$ and bitline $b_1$.  Rest of the devices are reset to 0 by using 0 as wordline input and 1 bitline input. Now, the third cube~$c_3$ can be computed, using steps identical to {\em Case 1} and the XOR can be performed with the result $c_1 \oplus c_2$. This process can be repeated till the entire ESOP has been computed. $\blacksquare$
\begin{figure}[t]
\centering
\scriptsize
\begin{tabular}{cc c ccc c ccc c ccc c cc}
\multicolumn{3}{c}{(a)} & & \multicolumn{3}{c}{(b)} & & \multicolumn{3}{c}{(c)} & & \\ \cline{2-3} \cline{6-7} \cline{10-11}
 & \multicolumn{1}{|c|}{$c_1$} & \multicolumn{1}{|c|}{$c_2$} & &
 & \multicolumn{1}{|c|}{$c_1$} & \multicolumn{1}{|c|}{$c_2$} & &
 0 & \multicolumn{1}{|c|}{$c_1$} & \multicolumn{1}{|c|}{$c_2$} & &\\  \cline{2-3} \cline{6-7} \cline{10-11}
  & \multicolumn{1}{|c|}{$0$} & \multicolumn{1}{|c|}{$0$} & & 
 1 & \multicolumn{1}{|c|}{$0$} & \multicolumn{1}{|c|}{$0$} & &
   & \multicolumn{1}{|c|}{$0$} & \multicolumn{1}{|c|}{$c_2$} & & 
\\\cline{2-3} \cline{6-7} \cline{10-11}
  & \multicolumn{1}{|c|}{$0$} & \multicolumn{1}{|c|}{$0$} & & 
 & \multicolumn{1}{|c|}{$0$} & \multicolumn{1}{|c|}{$\overline{c_2}$} & &
  & \multicolumn{1}{|c|}{$0$} & \multicolumn{1}{|c|}{$\overline{c_2}$} & &
\\\cline{2-3} \cline{6-7} \cline{10-11}
  & & & ~&
 & & $\overline{c_2}$ & & 
  & & $c_1$ & & 
\\
\multicolumn{3}{c}{(d)} & & \multicolumn{3}{c}{(e)} & & \multicolumn{3}{c}{(f)} & & \\\cline{2-3} \cline{6-7} \cline{10-11}
 1  & \multicolumn{1}{|c|}{$c_1$} & \multicolumn{1}{|c|}{$c_2.\overline{c_1}$} & & 
   & \multicolumn{1}{|c|}{$c_1$} & \multicolumn{1}{|c|}{$c_2 \oplus {c_1}$} & &
    & \multicolumn{1}{|c|}{$0$} & \multicolumn{1}{|c|}{$c_2 \oplus {c_1}$} & &\\\cline{2-3} \cline{6-7} \cline{10-11}
    & \multicolumn{1}{|c|}{$0$} & \multicolumn{1}{|c|}{$c_2+\overline{c_1}$} & & 
    & \multicolumn{1}{|c|}{$0$} & \multicolumn{1}{|c|}{$c_2+\overline{c_1}$} & &
    & \multicolumn{1}{|c|}{$0$} & \multicolumn{1}{|c|}{$0$} & &\\\cline{2-3} \cline{6-7} \cline{10-11}
        & \multicolumn{1}{|c|}{$0$} & \multicolumn{1}{|c|}{$\overline{c_2}$} & &
         & \multicolumn{1}{|c|}{$0$} & \multicolumn{1}{|c|}{$\overline{c_2}$} & &
         & \multicolumn{1}{|c|}{$0$} & \multicolumn{1}{|c|}{$0$} & &\\\cline{2-3} \cline{6-7} \cline{10-11}
   & &  $c_2+\overline{c_1}$        \\
\end{tabular}
\caption{\em ESOP computation on a $3\times2$ crossbar. (a) Cubes are computed on devices at wordline $e_2$. (b-e) Some of the intermediate steps of computing XOR of the two cubes are shown. (f) All devices, except the device holding the XOR of the cubes is reset to 0. If the ESOP has more cubes, the next cube $c_3$ would be computed at wordline~$e_2$ and bitline~$b_0$ and XOR would be computed for $c_3$ and $c_2 \oplus c_1$, followed by reset. This process is repeated till the entire ESOP has been computed.}
\label{fig:proofcube}
 \vspace{-0.6cm}
\end{figure}

The theorem~\ref{thm:32} guarantees that any ESOP can be computed in a crossbar with three wordlines and two bitlines. If the number of bitlines is greater, it is possible to reduce the delay by parallising operations. Boolean AND of two literals $a$ and $b$ can be expressed as $M_3(a,0,b)$.  $0$ can be used a common wordline input during computation of cubes in parallel feasible. Fig~\ref{fig:cubecomp} shows the computation of the cubes of an ESOP. Due to the crossbar constraints, all the  bitline-applied literals must be either available via the PIR or DMR simultaneously. 
This implies that all the applied literals either have to be primary inputs or must reside on the same wordline for parallel computation of the cubes.

At the end of completion of computation of the cubes, the cubes have to be XORed. 
Each XOR can be performed using that steps similar to the example shown in Fig.~\ref{fig:2inp}. Multiple XORs can be performed by means of a XOR reduction tree with logarithmic depth, in the number of terms to be XORed. In Fig.~\ref{fig:xorred}, there are four terms $x_i$ to be XORed. The XOR of $x_1$ and $x_2$ can proceed in parallel with the XOR of $x_3$ and $x_4$. Thereafter, the results $x_{12}$ and $x_{34}$ are XORed. It might happen that the numyclesber of cubes in an ESOP is greater the number of available bitlines in the crossbar. In that case, the computation of the cubes, followed by XOR reduction has to be iterated. 

\begin{figure}[t]
\centering
\def\arraystretch{1.5}
\begin{tabular}{lllllllllllllll}\cline{2-3} \cline{5-6} \cline{8-9} \cline{11-12}
\multicolumn{1}{l|}{1} & \multicolumn{1}{l|}{0} & \multicolumn{1}{l|}{0} &   \multicolumn{1}{l|}{0} & \multicolumn{1}{l|}{a} & \multicolumn{1}{l|}{$\overline{a}$} & \multicolumn{1}{l|}{0} & \multicolumn{1}{l|}{$a\overline{b}$} & \multicolumn{1}{l|}{$\overline{a}b$} & \multicolumn{1}{l|}{} & \multicolumn{1}{l|}{$a\overline{b}c$} & \multicolumn{1}{l|}{$\overline{a}bc$} &  &  \\ \cline{2-3} \cline{5-6} \cline{8-9} \cline{11-12}
                       & $\overline{a}$         & $a$                      &                          & $b$                      & $\overline{b}$                      &                        & $\overline{c}$                     & $\overline{c}$                     &                       &                                       &                                       &  & 
\end{tabular}
\caption{\em Computation of cubes of the ESOP $\overline{a}bc \oplus a\overline{b}c$.}
\label{fig:cubecomp}
\end{figure}
\begin{figure}[t]
	\centering
	\includegraphics[width=1.8in]{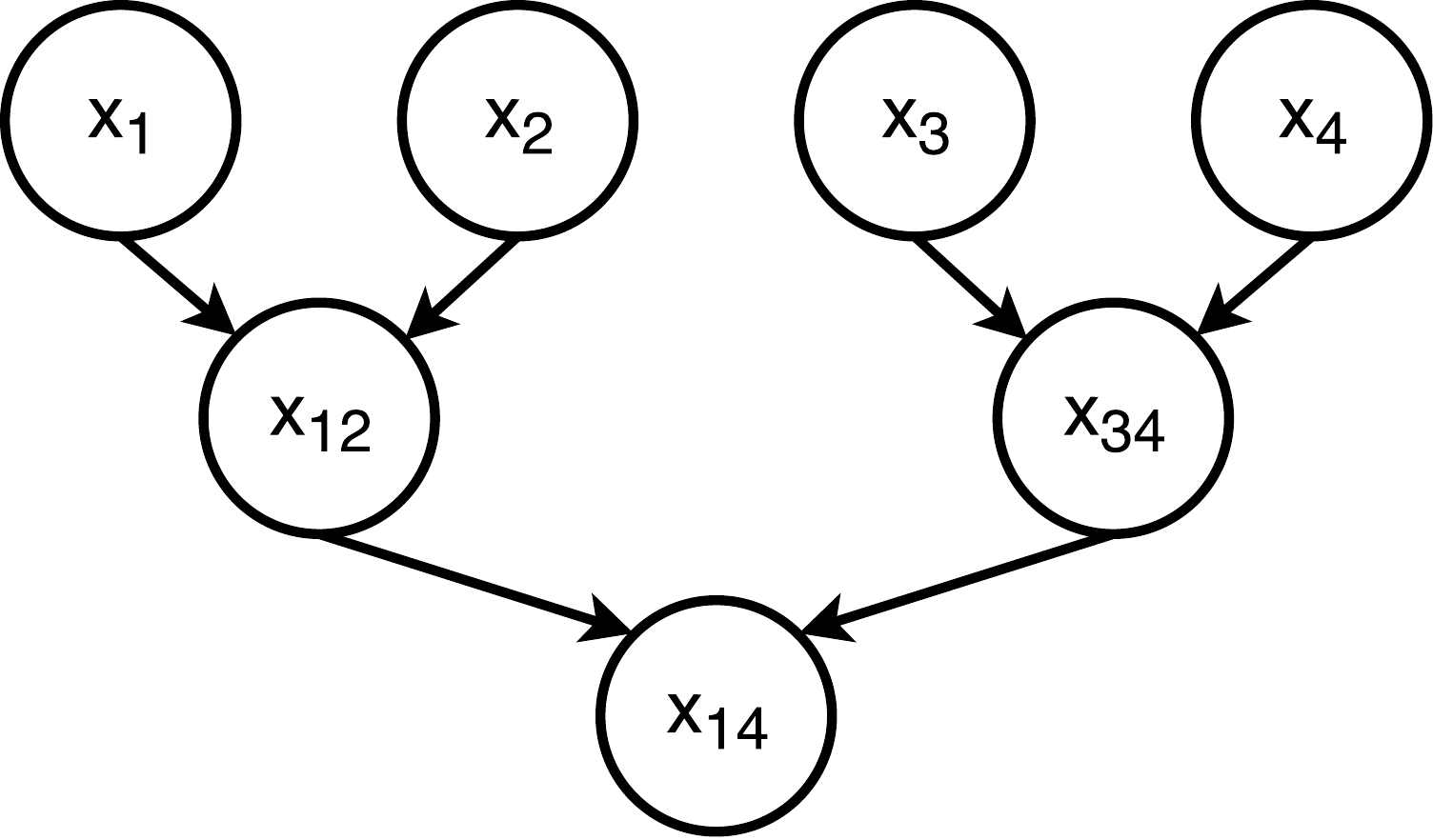}
	\caption{\em A XOR-reduction tree 4 terms $x_1, ..., x_4$. \mbox{$x_{12} = x_1 \oplus x_2$}~, $x_{34} = x_3 + x_4$ and \mbox{$x_{14} = x_{12}\oplus x_{34}$}}
	\label{fig:xorred}
	 \vspace{-0.8cm}
\end{figure}

\begin{algorithm}
\scriptsize
 \KwData{$E$, $CrossbarState$, $Loc$}
$result$ = $null$\;
\Do{Some cube has not been computed}
{
    $activeCubes$ = set()\;
    $currentLiterals$ = list()\;
    $v$ = max$_{l \in C}$ occ($l$)\;
    $currentLiteral$.append($v$)\;
    $activeCubes$.add(cube($v$));
    
    \While{ occ($currentLiteral$) $\le w_D$}
    {
	$allowedLiterals$ = findAllowed($E$,$currentLiterals$)\;
	Choose $v' \in allowedLiterals$ with max occurence in $cube(v') - activeCubes$ and occ($currentLiteral+v') \le w_D$;
	$currentLiteral$.append($v'$)\;
        $activeCubes$.add(cube($v'$));
    }
    And($currentLiterals$)\;
    \Do{All the activeCubes have not been computed}
    {
      $activeCubes'$ = set()\;
      $currentLiterals'$ = list()\;
      $v$ = max$_{l \in C}$ occ($l$,$activeCubes$)\;
      \While{ occ($currentLiteral$) $\le w_D$}
    {
	$allowedLiterals$ = findAllowed($E$,$currentLiterals$)\;
	Choose $v' \in allowedLiterals$ with max occurence in $cube(v') - activeCubes$ and occ($currentLiteral+v') \le w_D$;
	$currentLiteral$.append($v'$)\;
        $activeCubes'$.add(cube($v'$));
    }
    And($currentLiterals$)\;
    }
    $result$ = XorReduction($activeCubes$, $result$)\;
}
\caption{ESOP computation}
 \label{algo:esop}
\end{algorithm}

 \noindent The technique for ESOP computation is presented in Algorithm~\ref{algo:esop}. Once the ESOP has been evaluated, the result is written back to the position in the working area, as determined by the scheduling algorithm. 

\noindent \textbf{Discussion:} The proposed approach provides a novel solution to the area-constrained technology mapping problem. The target Boolean function is represented as an AIG, followed by partitioning into $k$-input LUTs and finally scheduling and computing these LUTs on the crossbar. The approach allows a feasible mapping for a variety of crossbar sizes, with some portion of the crossbar reserved for computation of ESOPs.

Instead of using AIGs for representing the functions, it is also feasible to represent the function using Majority Inverter Graph~(MIG). The native function realized by 
ReRAM devices is Boolean Majority three~($M_3$) with an input inverted. Therefore, MIGs have been used heavily in synthesis~\cite{shirinzadeh2016fast,techaware} and technology mapping~\cite{bhattacharjee2016delay} flows for ReRAM crossbar array. In the next section, we discuss another approach to the technology mapping problem using MIGs for logic
representation and constrained by only the word length of the DCM 
with focus on reducing the delay of mapping.
  \section{Delay-Constrained technology mapping}\label{sec:delay}
\noindent In this section, we present a method to generate instructions for the ReVAMP architecture that is focused at reducing the delay of mapping without constraints on the number of words required for mapping. In this method, we still consider the constraint on word length~$w_D$ during mapping. 

%

\subsection{Assign Host and Inputs to Nodes}
\noindent A ReRAM device has an internal state $Z$, and two inputlines---the wordline and bitline. A computation on it updates its internal state $Z$, in effect 
making the device the $host$ for the computation. For each internal node in an MIG, one of its parents hosts the computation and the remaining parents act as wordline and bitline inputs.
The computation of multiple independent nodes can be grouped into an Apply instruction if they have a common wordline input. Based on this,
we present a few rules to assign the host and the inputs of the nodes of an MIG.
\begin{itemize}
 \item If a node has multiple children in the same level, then it can be used as common wordline input for computing  those nodes.
 For instance, in   Fig.~\ref{fig:mig}, input $b$ can be used as common wordline input to compute $S_1$ and  $S_2$.
\item If an incoming edge to a node is marked inverted, then the corresponding parent can be used as the bitline input. 
In   Fig.~\ref{fig:mig}, $c$ and $S_2$ are used as bitline inputs to compute $S_1$ and $S_3$ respectively.
\item If there are no inverted incoming edges to a node, then a negated parent is used as input to that node. For node $S_2$ in   Fig.~\ref{fig:mig}, input $\overline{c}$ is used as bitline input.
\item The remaining parent is used as host for the node. The nodes $a$ and $S_1$ act as $host$ to compute $S_1$ and $S_3$ respectively in   Fig.~\ref{fig:mig}.
\end{itemize}
These rules ensure that the nodes with common inputs can share wordline inputs which is used for scheduling computation. We mark these assignments on the edges of the MIG, as shown in Fig.~\ref{fig:mig}.


\subsection{Group Nodes to Blocks} 
\noindent To compute an internal node in a MIG, we need to read out the wordline and bitlines inputs of the node and then apply these inputs to the host. Given  that only a single word can be read out in a clock cycle, the wordline and bitline inputs of the node must reside on the same wordline to allow efficient computation of the node. This creates a constraint
that for each node in an MIG --- the wordline and the bitline inputs should be placed in the same word. We call this grouping a \textit{block}.  
  
Further, as read-outs are non-destructive, blocks can be merged if they have common inputlines. This reduces the number of devices required, with the merged block having only one copy of the common inputline. Note that  blocks can be merged only if the number of inputs in the resultant block does not exceed the word length.

Also, a pair of blocks in the same level that have hosts which share a wordline input should be merged. This host-based merge along with  merge of the corresponding blocks with the inputlines of these hosts permits computation of the nodes in the same level with shared wordline in a single cycle, thereby reducing delay.

\begin{algorithm}
\scriptsize
 \KwData{G, pi, po}
 \KwResult{blockList}
 blockCount = 0\;
 \For{$node_{out}~\in~po$}
 {
      addBlock([($node_{out}.host)$],blockCount)\;
      addBlock([($node_{out}.wl,node_{out}.bl)$],blockCount)\;
      addInversionBlock($el$)\;
 }
 mergeBlock()\;
 \For{$l = level_{max}$; $l > 0$; $l=l-1$}
 {
    \For{$block \in blockList$}
 {
	    \For{$el \in block$}
	{
	    \If{$el \notin pi$ and $el.level == l$}
	    {
	       replace($el$, $el.host$)\;
	       addBlock([$el.wl$, $el.bl$], blockCount)\;
	       addInversionBlock($el$)\;
	    }
	    
	}
 }
 mergeBlock()\;
 }
\caption{Block Formation Algorithm}
\label{algo:block_merge}
\end{algorithm}
The algorithm of the block formation is shown in Algorithm~\ref{algo:block_merge}. The lines 2-5 creates the blocks considering the placement constraint on the input lines of the output nodes. 
The addInversionBlock method adds the positive nodes as blocks to the blockList, if the added blocks have inverted values. Only a single positive node is added to blockList, corresponding
to multiple copies of a  negated node. The mergeBlock method merges blocks based on the input line and host based merge constraints. The replace method replaces a node in a block with its host node. 
\begin{table}[ht]
\caption{\em BlockList update with BlockMerge algorithm for MIG of Fig.~\ref{fig:mig}}
\label{table:migblock}
 \begin{tabular}{|r|l|}\hline
 \textbf{Level} & \textbf{BlockList} \\ \hline
 Output & [[1,$S_4$)]] \\ \hline
 3  & [[(1,$S_3$,h)],[(2,$d$,i),(2,$\overline{e}$,i)]] \\ \hline
 2 & [[(1,$S_1$,h)],[(2,$d$,i),(2,$\overline{e}$,i)],[(3,$c$,i),(3,$S_2$,i)]] \\ \hline
 1 & [[1:($a$,h)],[2:($d$,i),($\overline{e}$,i)],[3:($c$,i),($a$,h)],[4:($b$,i),($c$,i)],[5:($b$,i),($\overline{c}$,i)]]  \\ 
 1 & [[1:($a$,h)],[2:($d$,i),($\overline{e}$,i)],[3:($c$,i),($a$,h)],[4:($b$,i),($c$,i),($\overline{c}$,i)]]  \\ 
 1 & [[1:($a$,h),($c$,i),($a$,h)],[2:($b$,i),($\overline{e}$,i)],[4:($b$,i),($c$,i),($\overline{c}$,i)] \\ \hline
 \end{tabular}
 \vspace{-0.8cm}
 \end{table}
\begin{example}For a word length~($w_D$) of 3, Table~\ref{table:migblock} shows	the working of the block formation algorithm on the MIG of Fig.~\ref{fig:mig}.
Starting at the output node, blockList has a single block. At level 3, node $S_4$ is replaced with its host and inputlines. Since these 
 two blocks do not have any common inputlines or hosts, they cannot be merged. At level 2, node $S_3$ gets replaced and the inputlines are 
added to a new block.  At level 1, nodes $S_1$ and $S_2$ are replaced by their hosts $a$, and the inputlines are inserted in two new blocks. Blocks 4 and 5 have a 
common inputline $b$ and are hence merged. Blocks 2 and 4 have common inputs, but cannot be merged as the length (four) of the resultant block will exceed the given word length.
Thereafter, since the two $a$ host nodes have the same wordline, blocks 1 and 3 get merged, but both copies of the host are retained, using the host-merge constraint.
\end{example}
\begin{figure*}[ht]
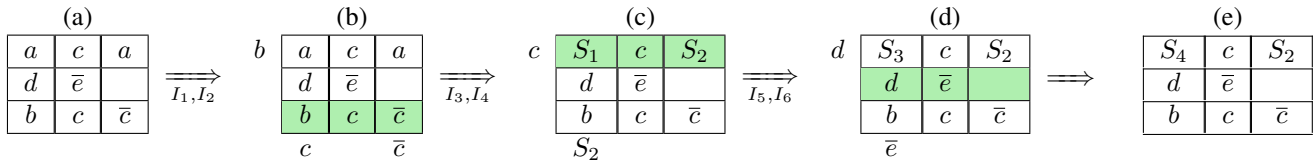

 \begin{tabular}{ccc cr ccc cr ccc cr ccc cc ccc} 
 \multicolumn{3}{c}{(a)} & & & \multicolumn{3}{c}{(b)} & & & \multicolumn{3}{c}{(c)} & & & \multicolumn{3}{c}{(d)} & & & \multicolumn{3}{c}{(e)}\\ \hhline{---~~---~~---~~---} \cline{21-23}
  \multicolumn{1}{|c|}{$a$} & \multicolumn{1}{|c|}{$c$} & \multicolumn{1}{|c|}{$a$} 
  & \multirow{ 3}{*}{$\xRightarrow[I_1, I_2]{\text{}}$} & $b$ & \multicolumn{1}{|c|}{$a$} & \multicolumn{1}{|c|}{$c$} & \multicolumn{1}{|c|}{$a$}       
  &  \multirow{ 3}{*}{$\xRightarrow[I_3, I_4]{\text{}}$} & $c$ & \multicolumn{1}{|c|}{\ccol $S_1$} & \multicolumn{1}{|c|}{\ccol $c$} & \multicolumn{1}{|c|}{\ccol $S_2$}   
  &  \multirow{ 3}{*}{$\xRightarrow[I_5, I_6]{\text{}}$} & $d$ & \multicolumn{1}{|c|}{ $S_3$} & \multicolumn{1}{|c|}{ $c$} & \multicolumn{1}{|c|}{ $S_2$}    
  &  \multirow{ 3}{*}{$\xRightarrow[~~~~]{\text{}}$}   & & \multicolumn{1}{|c|}{ $S_4$} & \multicolumn{1}{|c|}{ $c$} & \multicolumn{1}{|c|}{ $S_2$}  
  \\ \hhline{---~~---~~---~~---} \cline{21-23}
  \multicolumn{1}{|c|}{$d$} & \multicolumn{1}{|c|}{$\overline{e}$} & \multicolumn{1}{|c|}{~}  
  & & &\multicolumn{1}{|c|}{$d$} & \multicolumn{1}{|c|}{$\overline{e}$} & \multicolumn{1}{|c|}{~}  
  & & &\multicolumn{1}{|c|}{$d$} & \multicolumn{1}{|c|}{$\overline{e}$} & \multicolumn{1}{|c|}{~}  
  & & &\multicolumn{1}{|c|}{\ccol $d$} & \multicolumn{1}{|c|}{\ccol $\overline{e}$} & \multicolumn{1}{|c|}{\ccol~}
 & & &\multicolumn{1}{|c|}{ $d$} & \multicolumn{1}{|c|}{ $\overline{e}$} & \multicolumn{1}{|c|}{~}
  \\ \hhline{---~~---~~---~~---} \cline{21-23}
  \multicolumn{1}{|c|}{$b$} & \multicolumn{1}{|c|}{$c$} & \multicolumn{1}{|c|}{$\overline{c}$}  
  & & & \multicolumn{1}{|c|}{\ccol $b$} & \multicolumn{1}{|c|}{\ccol $c$} & \multicolumn{1}{|c|}{\ccol $\overline{c}$}  
  & & & \multicolumn{1}{|c|}{$b$} & \multicolumn{1}{|c|}{$c$} & \multicolumn{1}{|c|}{$\overline{c}$} 
  & & & \multicolumn{1}{|c|}{$b$} & \multicolumn{1}{|c|}{$c$} & \multicolumn{1}{|c|}{$\overline{c}$} 
  & & & \multicolumn{1}{|c|}{$b$} & \multicolumn{1}{|c|}{$c$} & \multicolumn{1}{|c|}{$\overline{c}$} 
  \\\hhline{---~~---~~---~~---} \cline{21-23}
  & & 
  & & & $c$ & & $\overline{c}$ 
    & & & $S_2$ & &
      & & & $\overline{e}$ & & 
 \end{tabular}
\caption{\em DCM state transition during computation. (a) DCM state after loading the primary inputs. (b-d) The intermediate DCM states during computation. (e) The final DCM state. The green coloured row represents the read out wordline.}
\label{fig:state}
 \vspace{-0.8cm}
\end{figure*}

\subsection{Pack Blocks in Words} 
\noindent At the end of scheduling computation, we have blocks of elements, which have to be placed in the same wordline. The number of elements in each block is less than or equal
to $w_D$, the number of bits in a word. Now, these blocks have to packed in the DCM using the minimum number of words. The problem can be formulated
as a bin packing problem  as defined below.
\begin{algorithm}
\scriptsize
 \KwData{blockList,wordlength}
 \KwResult{blockToWord}
 wordToBlock = HashMap()\;
 wc = 0\;
 \For{$block~\in~blockList$}
 {
  $assigned$ = False\;
    \For{$w~\in~wordToBlock$}{
      \If{$wordToBlock$[$w$].occupied + $block$.length $<$ $wordlength$}
      {
	 $wordToBlock$[$w$].occupied =$wordToBlock$[$w$].occupied + $block$.length\;
	 $wordToBlock$[$w$].append(block)\;
	 $assigned$ = True\;
      }
      \If{$assigned$ == False}
      {
	$wc$ = $wc$+$1$\;
	$wordToBlock$[$wc$].append(block)
	$wordToBlock$[$wc$].occupied = $block$.length\;
      }
   }
 }
 \caption{First-fit Algorithm}
 \label{label:ff}
\end{algorithm}

 Consider each word in the DMR as a bin, with capacity $w_D$. Each block $b_i$ has a value $v_i$, $v_i > 0$. Each block must be assigned to a bin such the 
 total value of the objects assigned to the bin is less than or equal to $w_D$. The objective is to minimize the number of bins required to assign all the block, without violating the capacity constraint.

This \textit{first-fit} algorithm provides a 2-factor approximation, i.e., the number of words required by the algorithm is at most twice the number of words required by the optimal solution.
\begin{example}
For the example, the blocks determined by the Block Formation algorithm are placed in a separate wordline, as shown in Fig.~\ref{fig:state}~(a).
\end{example}

\subsection{Generation and Scheduling instructions} 
\noindent The primary inputs have to be loaded into the DCM before computation of the internal nodes of the MIG can begin. In each clock cycle, $w_D$ primary inputs can be read. 
The primary inputs are loaded via the bitline and hence the inverted values are stored in a single clock cycle. To store non-inverted primary inputs,
the primary inputs are written to a wordline, thereby storing it in inverted form. Then, the inverted value is read out and applied
via the bitline to store the non-inverted value to the required wordline. A single extra wordline is used for storage of the intermediate inverted primary input, and this wordline 
is reset, after each use.

All the nodes in level $i$ are scheduled for computation before any node at  level $i+1$ is scheduled. The nodes in the same level can be scheduled in any order as they do not have
any data dependencies. The nodes in a level with hosts of the which are in the same block, and the corresponding inputlines are also placed together in the same block, are scheduled 
for computation together. Once all the nodes in a level have been computed,  we determine whether any inverted copies of the nodes are required for computation of nodes present at a higher level. If inverted copies are needed, the node is read out and stored in inverted form in the required block by writing through the bitline. Each computation is expressed as an {\em Apply} instruction
and read operations are expressed as {\em Read} instructions.

\begin{table}[ht]
\vspace{0.2cm}
\centering
\caption{\em Instruction sequence to compute MIG in Fig.~\ref{fig:mig}.}
\label{table:ins}
\begin{tabular}{rl}\bottomrule
& \textbf{Instruction} \\ \midrule
$I_1$ & Read 0 \\ 
$I_2$ & Apply 2 11 0~~1 1~~0 0~~1 2 \\ 
$I_3$ & Read 2 \\ 
$I_4$ & Apply 2 11 1~~1 2~~0 0~~0 0\\
$I_5$ & Read  1 \\ 
$I_6$ & Apply  2 11 0~~1 1~~0 0~~0 0 \\ \toprule
\end{tabular}
 \vspace{-0.8cm}
\end{table}

\begin{example}
 Table~\ref{table:ins} shows the sequence of instructions used to compute the example MIG, and Fig.~\ref{fig:state} shows the changes in DCM state on application of the \textbf{Apply} instructions.  Note that the additional instructions needed to initialize the DCM are not shown. The inputs 
to compute nodes $S_1$ and $S_2$ are in word 0 and are read out. The hosts of nodes $S_1$ and $S_2$ are in word 2, and therefore $I_2$ computes these nodes in word 2.
The inputs to compute $S_3$ are in word 2, and are read out by $I_3$. $I_4$ computes $S_3$ in host $S_1$. 
Finally to compute $S_4$,  $I_5$ reads out word 1 and  $I_5$ applies the required inputs to $S_3$. 
\end{example}

\noindent\edit{ \textbf{Discussion:} Even though the two approaches have been discussed with AIG and MIG as the input data structures, the data structures can be used interchangeably. To use MIG in the area-constrained mapping approach, the MIG can be directly partitioned into LUTs and the rest of the mapping flow can be used. Similarly, the AIG can be converted to an MIG by introducing constant `0' as the third input to each node, and the rest of the delay-constrained mapping flow can be used. Due to the inherent sequential nature of computation on  and the crossbar constraints, employing traditional synthesis optimization techniques, such as depth reduction, do not directly translate into lower delay after technology mapping. However, it is possible to make the synthesis optimization technique technology-aware to aid the technology mapping flow, as demonstrated recently by Bhattacharjee et al.~\cite{bhattacharjee2018technology}. } 

  \section{Experimental results}\label{sec:experimental}
\noindent  We have implemented the  proposed compilation flow for the ReVAMP architecture using Python. 
The algorithm was evaluated using the EPFL benchmarks\footnote{http://lsi.epfl.ch/benchmark}. For area-constrained mapping, we used ABC for generating the initial AIG and also for ESOP expansion~\cite{abcalan}. 
Each run is limited to 2 hours, exceeding which the program is terminated. The major amount of time in mapping is spent in ESOP expansion.

\begin{figure*}[ht]
\begin{adjustbox}{minipage=\linewidth,scale=0.9}
	\begin{subfigure}[t]{3.8in}
		\caption{ ~~}
		\label{fig:ac97delay}
		\includegraphics[width = 0.96\linewidth]{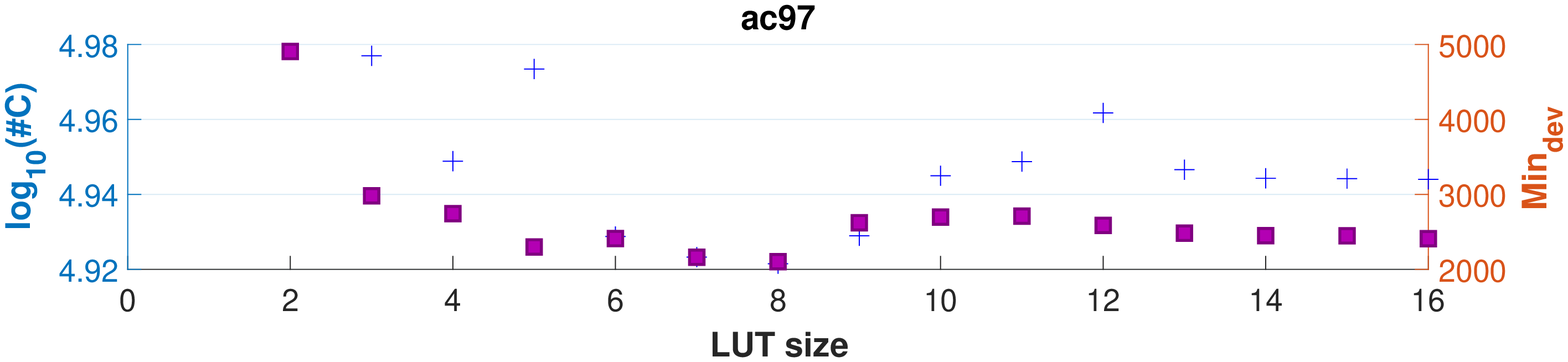}
	\end{subfigure}
	\begin{subfigure}[t]{3.8in}
		\caption{ ~~}
		\label{fig:i2cdelay}
		\includegraphics[width = 0.96\linewidth]{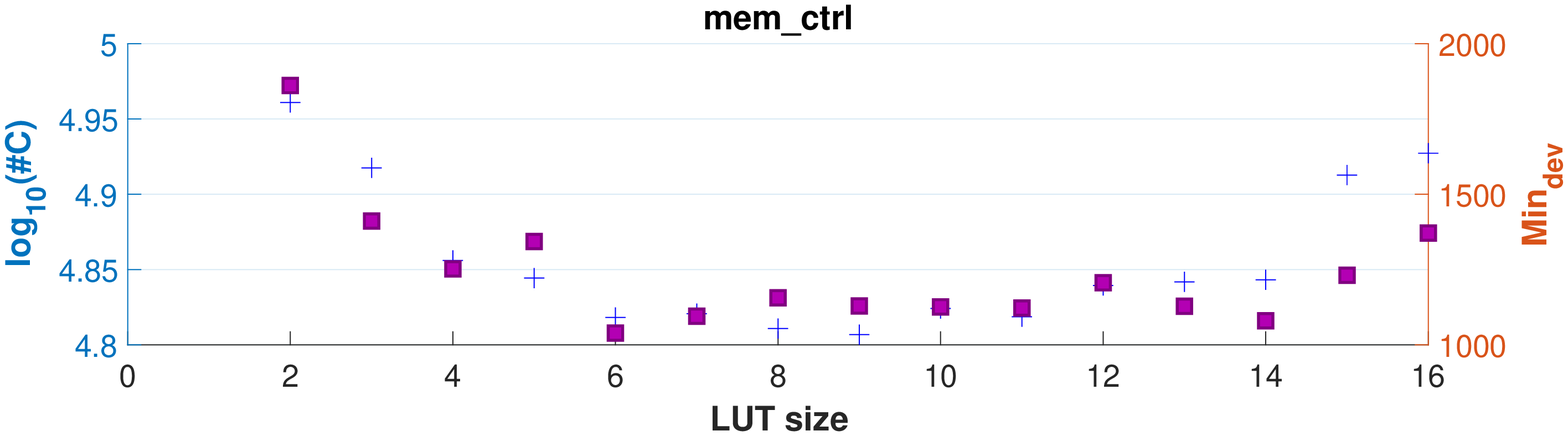}
	\end{subfigure}\\
    \begin{subfigure}[t]{3.8in}
    	\caption{ ~~}
    	\label{fig:mul32delay}
    	\includegraphics[width = 0.96\linewidth]{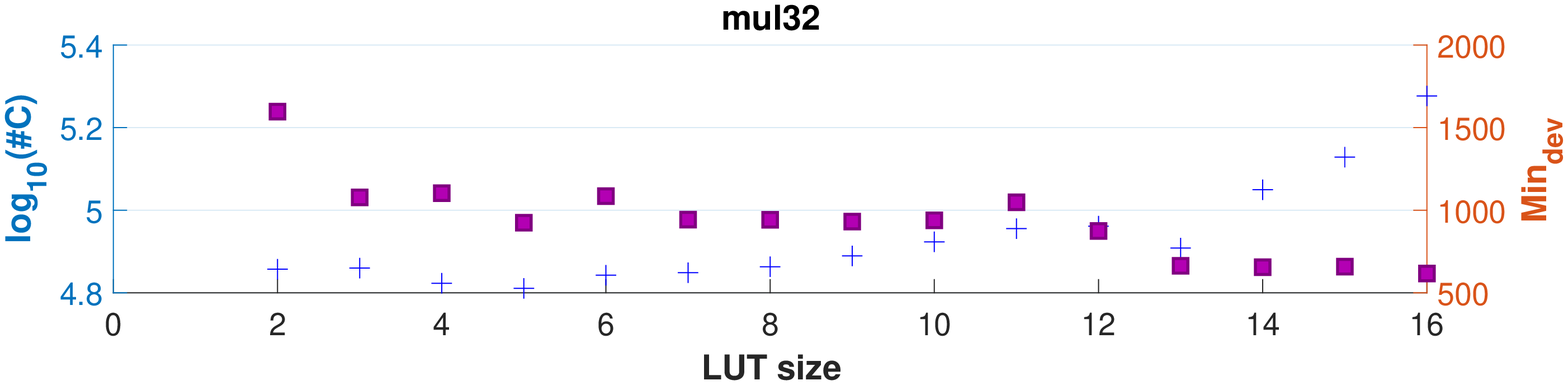}
    \end{subfigure}
    \begin{subfigure}[t]{3.8in}
    	\caption{ ~~}
    	\label{fig:revxdelay}
    	\includegraphics[width = 0.96\linewidth]{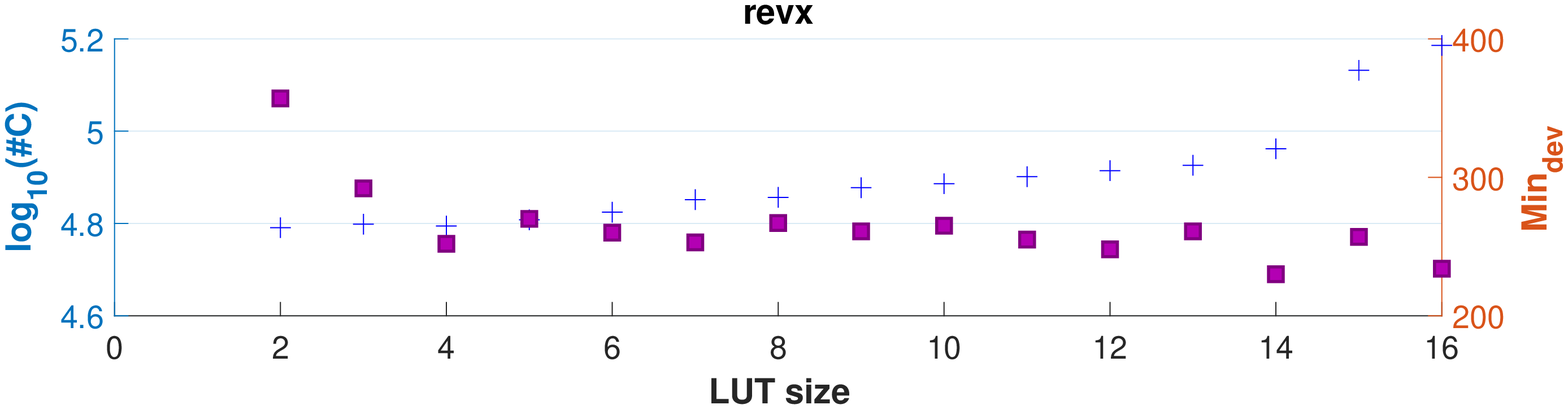}
    \end{subfigure}
 
	\caption{\em Impact of LUT size (k) on delay (\#C) for crossbar dimensions $64\times64$. The blue + symbol indicates the delay of mapping $\#C$ in the log scale. The violet squares denote the minimum number of devices $Min_{dev}$ for a feasible mapping.}
	\label{fig:delayk}
	  \end{adjustbox}
	
\end{figure*}

For all the EPFL benchmarks, Table~\ref{table:benchmarks_results} presents the results of the area-constrained mapping for varying number of LUT inputs $k$ for fixed crossbar dimension of $64 \times 64$.
With increase in $k$, the number of LUTs~($\#N_{LUT}$) in the LUT graph reduces, along with reduction in the number of levels~($\#L$). For the given crossbar dimensions, 61 words are available for storing the intermediate results and 3 are reserved for computing the ESOPs. Some of the  benchmarks could not be mapped~(marked by $\times\times$) due to violation of the feasibility criteria~($Min_{Dev} > 3904$), presented in Equation~(\ref{eq:mindev}). 

To analyze the impact of increasing number of LUT inputs~($k$) on delay and $Min_{Dev}$ in detail, we consider four large benchmarks from the EPFL benchmark suite for crossbar dimension $64\times 64$. The results are shown in Fig.~\ref{fig:delayk}. The effect of $k$ on $Min_{dev}$ is dependent on the benchmark itself. For example, with increase in value of $k$,  $Min_{dev}$ for the benchmark {\em mul32}  decreases but for {\em mem\_ctrl}, $Min_{dev}$ increases for larger values of $k$, as evident from the Fig.~\ref{fig:delayk}. The delay of the mapping~(in terms of number of cycles $\#C$) closely follows the trend of $Min_{dev}$ i.e. with increase in $Min_{dev}$, the overall number of cycles required for mapping increases. This is because with increase in $Min_{dev}$, less number of crossbar devices can be reset at any given time, which leads to reduction in the parallelization of operations during the ESOP computation. For the benchmark $mul32$, notice the sharp rise in delay of mapping on changing $k$ from 13 to 16. The number of cubes in the ESOP expression for increased consistently, resulting in the increased time of computation of the ESOP expression, that increases the overall delay of mapping. Also, for large values of $k$~($k \ge 28$), the time required for each ESOP expansion increases considerably~($ > 2$ hours), which leads to long execution time for mapping an entire benchmark. 

\begin{figure}[th]
       \includegraphics[width =\linewidth]{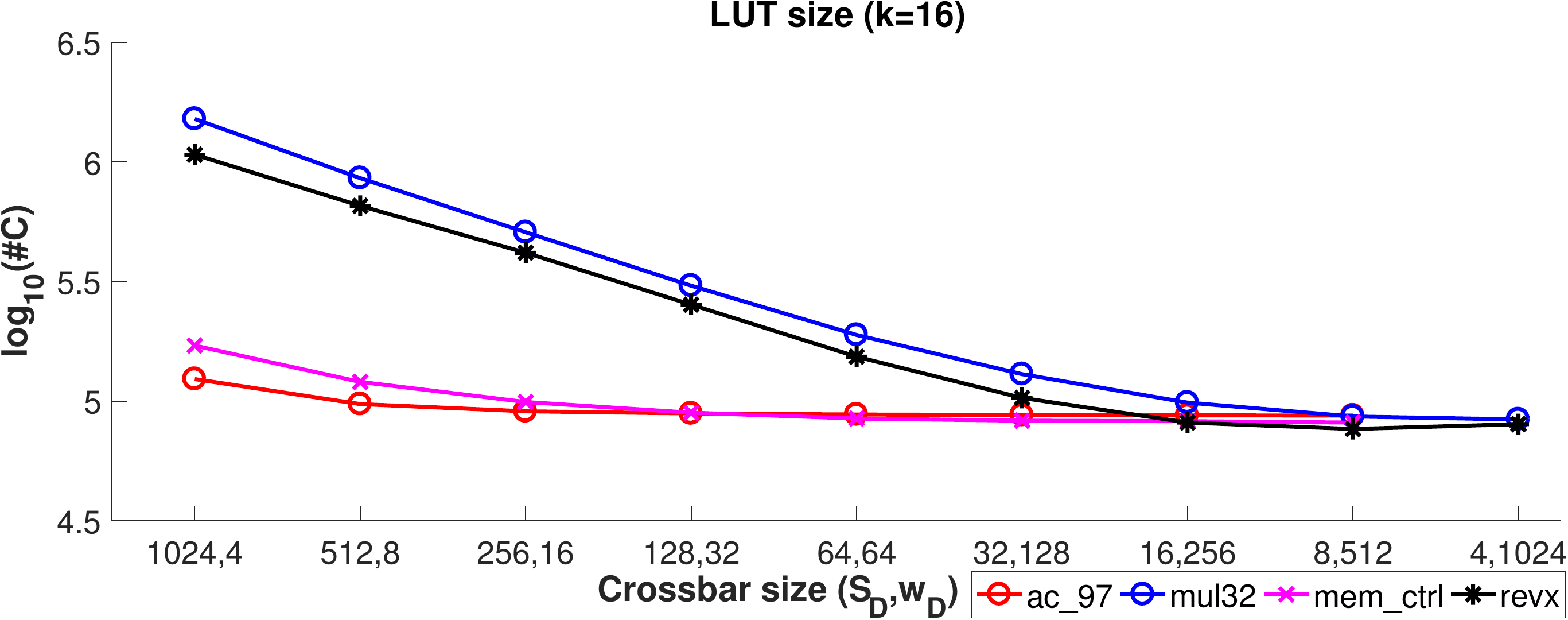}
       \caption{\em Impact of crossbar dimensions on delay~($\#C$) of area-constrained mapping. 4096 devices are used for all the mappings and the number of bitlines~($w_D$) is increased from 4~to~1024. }
       \label{fig:crossbardelay}
\end{figure}

\begin{table*}[ht]
	\caption{{\em Performance of area-constrained mapping on crossbar size ( $S_D\times w_D$ )=($64 \times 64$). $\times\times$ denote the benchmarks that cannot be mapped with the current crossbar due to $Min_{Dev}>3904$.} -- -- {\em indicate the benchmarks that did not complete mapping within the time limit~(2 hours).}}
	\label{table:benchmarks_results}
{\scriptsize
	\begin{tabular}{lrrrr r rrrr r rrrr r }\toprule
	\multirow{2}{*}{$Benchmark$} & \multicolumn{4}{c}{$k=4$} & & \multicolumn{4}{c}{$k=8$} & & \multicolumn{4}{c}{$k=16$}  \\
	\cmidrule{2-5}  \cmidrule{7-10}  \cmidrule{12-15}
	& 	 {$\#N_{LUT}$} & {$\#L$} & {$Min_{Dev}$} & {$\#C$} &	&  {$\#N_{LUT}$} & {$\#L$} & {$Min_{Dev}$} & {$\#C$} &  & {$\#N_{LUT}$} & {$\#L$} & {$Min_{Dev}$} & {$\#C$}  \\
	\midrule

ac97\_ctrl.v & 3900 & 4 & 2743 & 88893 &  & 2630 & 3 & 2102 & 83464 &  & 2409 & 2 & 2409 & 87908  \\
aes\_core.v & 8978 & 8 & 7006 & $\times\times$ &  & 902 & 3 & 890 & 33804 &  & 816 & 2 & 816 & 40091 \\
comp.v & 8918 & 50 & 2624 & 188156 &  & 5695 & 29 & 1987 & 179905 &  & 4169 & 19 & 2143 & 301379 \\
des\_area.v & 2020 & 10 & 1085 & 44303 &  & 699 & 5 & 505 & 29885 &  & 884 & 3 & 859 & 57017 \\
des\_perf.v & 34260 & 6 & 26309 & $\times\times$ &  & 11500 & 3 & 11476 & $\times\times$ &  & 9713 & 2 & 9713 & $\times\times$  \\
diffeq1.v & 6652 & 72 & 1736 & 164355 &  & 3949 & 38 & 1247 & 169057 &  & 2710 & 22 & 1079 & 414193  \\
div16.v & 1033 & 85 & 151 & 23557 &  & 710 & 39 & 131 & 27276 &  & 507 & 20 & 155 & 51407  \\
DSP.v & 14679 & 22 & 4513 & $\times\times$ &  & 9025 & 11 & 3823 & 323789 &  & 7195 & 7 & 3733 & 416307 \\
ethernet.v & 20178 & 10 & 11855 & $\times\times$ &  & 14385 & 6 & 10462 & $\times\times$ &  & 12113 & 3 & 9869 & $\times\times$  \\
hamming.v & 696 & 24 & 243 & 16676 &  & 498 & 13 & 221 & 19841 &  & 413 & 8 & 220 & -- --\\
i2c.v & 347 & 7 & 255 & 7264 &  & 219 & 3 & 157 & 5929 &  & 156 & 2 & 156 & 5921  \\
log2.v & 10331 & 107 & 2996 & 257920 &  & 4415 & 48 & 1208 & 240998 &  & 2859 & 25  & 723 &  -- --  \\
MAC32.v & 2828 & 29 & 1347 & 75746 &  & 1624 & 13 & 979 & 72908 &  & 967 & 7 & 775 & 110020 \\
max.v & 1023 & 54 & 735 & 25815 &  & 687 & 24 & 658 & 30267 &  & 615 & 12 & 615 & 45412  \\
mem\_ctrl.v & 3523 & 15 & 1252 & 71790 &  & 2372 & 7 & 1156 & 64689 &  & 2038 & 4 & 1371 & 84581 \\
MUL32.v & 2458 & 18 & 1103 & 66538 &  & 1580 & 9 & 942 & 72954 &  & 879 & 6 & 617 & 189105  \\
mult64.v & 7438 & 87 & 3105 & 229381 &  & 3937 & 40 & 1418 & 208960 &  & 2860 & 20 & 1109 & 468718  \\
pci\_bridge32.v & 6384 & 10 & 2435 & 143697 &  & 4585 & 6 & 2294 & 142557 &  & 4189 & 3 & 2927 &  -- --  \\
pci\_spoci\_ctrl.v & 373 & 7 & 296 & 7294 &  & 242 & 4 & 197 & 7144 &  & 119 & 3 & 96 & 4859 \\
revx.v & 2659 & 61 & 252 & 62304 &  & 1644 & 34 & 267 & 71856 &  & 559 & 15 & 234 & 153372  \\
sasc.v & 207 & 3 & 161 & 4876 &  & 147 & 2 & 147 & 4411 &  & 132 & 1 & 132 & 5038  \\
simple\_spi.v & 279 & 6 & 182 & 6002 &  & 187 & 3 & 137 & 5641 &  & 150 & 2 & 150 & 5762 \\
spi.v & 1211 & 11 & 656 & 27253 &  & 906 & 5 & 692 & 30900 &  & 491 & 3 & 359 & 36109 \\
sqrt32.v & 264 & 112 & 83 & 8103 &  & 206 & 37 & 79 & 10899 &  & 158 & 15 & 70 & 33350 \\
square.v & 5721 & 83 & 3128 & 154151 &  & 3466 & 36 & 1894 & 131707 &  & 2557 & 17 & 1672 & 193112  \\
ss\_pcm.v & 127 & 3 & 109 & 3040 &  & 100 & 2 & 100 & 3281 &  & 98 & 1 & 98 & 3243  \\
systemcaes.v & 3255 & 11 & 1830 & 79168 &  & 1565 & 6 & 693 & 59813 &  & 1033 & 4 & 692 & 49199  \\
systemcdes.v & 1109 & 9 & 576 & 26040 &  & 516 & 3 & 497 & 15857 &  & 290 & 2 & 290 & 19303  \\
tv80.v & 2822 & 19 & 1150 & 64901 &  & 1523 & 10 & 802 & 54089 &  & 769 & 5 & 562 & 49165  \\
usb\_funct.v & 4678 & 12 & 2330 & 110240 &  & 2765 & 6 & 1383 & 91873 &  & 2177 & 3 & 1455 & 126515  \\
usb\_phy.v & 181 & 4 & 151 & 3382 &  & 120 & 2 & 120 & 3061 &  & 111 & 1 & 111 & 3046  \\

		\bottomrule
	\end{tabular}
}
 \vspace{-0.8cm}
\end{table*}

 We analyze the impact of crossbar dimension on the delay of mapping. Keeping the overall 
 number of devices fixed to $4096$, we vary the number of bitlines from $4$~to~$1024$. The results
 are shown in Fig.~\ref{fig:crossbardelay}. With increase in the number of bitlines, the crossbar permits greater number of parallel operations that can be carried out in a word. This parallelism is harnessed by the ESOP computation technique, which leads to reduction in delay of mapping for the entire benchmark.


%

\newcommand{\ra}[1]{\renewcommand{\arraystretch}{#1}}


\begin{figure}[ht]
	\begin{subfigure}[t]{3.8in}
	\caption{~~}
	\label{fig:delay}
	\includegraphics[width=3.4in]{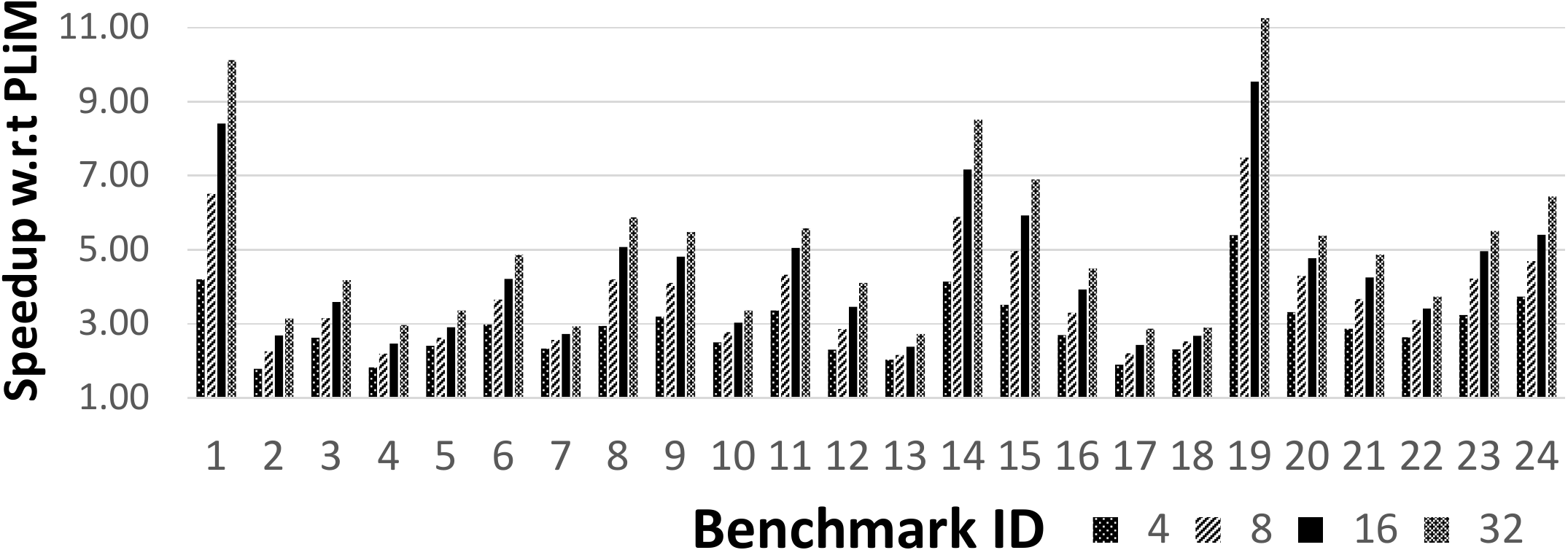}
	\end{subfigure}\\
	\begin{subfigure}[t]{3.8in}
	\caption{ ~~}
	\label{fig:util}
	\includegraphics[width=3.5in]{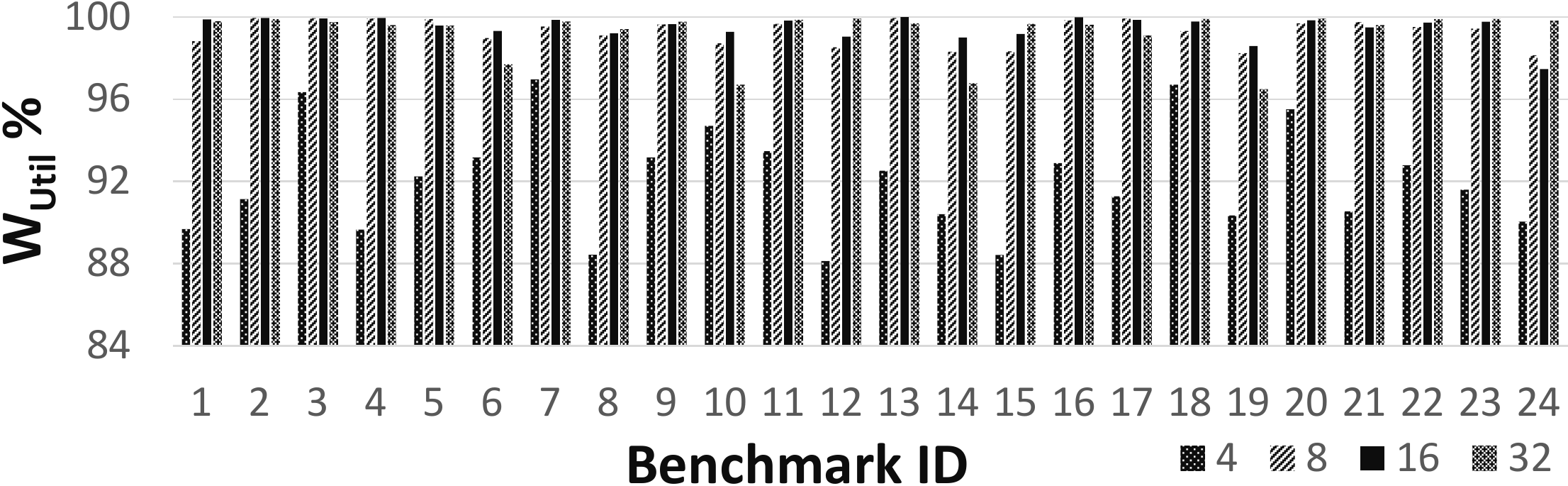}
	\end{subfigure}
	\caption{\em For varying word length $w_D=\{4,8,16,32\}$ (a)~Speedup achieved by the delay constrained mapping on the ReVAMP architecture against PLiM.  (b)~Word utilization achieved by the delay constrained mapping.}
\end{figure}

The results of delay constrained mapping are presented in Table~\ref{table:results} for word length~($w_D$) of 16-bits.  For most of the benchmarks, the compilation time to generate the instructions,
was a few seconds while for the larger benchmarks, the compilation process finished under 20 minutes. 
The number of Read and Apply instructions are shown in column $I_A$ and $I_R$ respectively while the total number of instructions is $I_{Total}$. The number of blocks created by mapping is $\#B$.
The total delay~($\#C$)  of the mapping solution is the number of cycles to
complete computation of the benchmark by the ReVAMP architecture.
\edit{Gaillardon et al.~\cite{Gaillardon} proposed the PLiM computer, 
which has a single instruction --- $RM_3~A,B,Z$. Assuming 16-bit words, each instruction results in the following micro operations on the memory array: Read @A~(32 bits), Read @B~(32 bits), Read @Z~(32 bits), Read~A ~(1 bit), Read B~(1 bit), Write @Z~(1 bit). This corresponds to 9 R/W~cycles on the considered machine. Therefore, minimum number of cycles $D_{P*}$ required by PLiM~\cite{Gaillardon} to compute any MIG is 9\#N, where \#N is the number of nodes in MIG. This delay does not include the additional delay required for computing the negated valued of the nodes. For each benchmark, $\#C$ is significantly
lower than the $D_{P*}$ achieved by the PLiM computer. This is a fair comparison since PLiM computer also used a word-length of $16$-bits. The ReVAMP architecture outperforms PLiM computer for the same 16-bit word by a factor of $4.38\times$ on average and $9.5\times$ at the maximum. For the ReVAMP architecture, on average, almost 30\% of the computation time is spent in computing negated value of the nodes. Thus, the ReVAMP architecture would further outperform the PLiM computer, when the actual number of cycles  required  by PLiM computer for computation with negations will be considered. Synthesis techniques can be used to reduce the number of nodes in the MIG for reducing the delay of executing a benchmkark on the PLiM, as suggested in~\cite{soeken2016mig}, but similar techniques can also be used for optimizing the input data structures of the proposed technology mapping flow~\cite{bhattacharjee2018technology}.} 

In Fig.~\ref{fig:delay}, the speed up achieved by the ReVAMP architecture against the PLiM computer is  presented for various word lengths.
Even for a small word length of 4, the ReVAMP architecture gains in performance over the PLiM architecture by a factor of $2.9\times$ on average.
This shows that harnessing the inherent parallelism of ReRAM crossbar arrays for computation provides considerable performance gains. This justifies the VLIW nature of the ReVAMP architecture and demonstrates the effectiveness of the delay constrained mapping.

\begin{table*}[ht]
\centering
\caption{\em Performance of the ReVAMP architecture on EPFL benchmarks for $w_D$=16. }
\label{table:results}
{
 \begin{tabular}{rlr rrr rrr rrr r rrrrrrr} \hline
ID &Benchmark& $N_{Maj}$ & $I_{A}$ & $I_R$ & $I_{total}$ &\#B & $S_D$ & $W_{Util}$ &  $\#C$ & $D_{P*}$ \\\hline
1&ac97\_ctrl&15253&8803&7520&16323&2330&933&99.88&16325&137277 \\
2&comp&18967&32297&31293&63590&3114&1182&99.96&63592&170703 \\
3&des\_area&4629&5971&5639&11610&1073&305&99.92&11612&41661 \\
4&div16&4440&8375&7825&16200&702&342&99.96&16202&39960 \\
5&hamming&2280&3603&3450&7053&623&180&99.58&7055&20520 \\
6&i2c&1263&1450&1247&2697&297&83&99.32&2699&11367 \\
7&MAC32&9489&15980&15363&31343&3045&784&99.85&31345&85401 \\
8&max&4854&4431&4184&8615&990&314&99.2&8617&43686 \\
9&mem\_ctrl&9569&9497&8405&17902&2032&625&99.65&17904&86121 \\
10&MUL32&9226&14047&13389&27436&2406&718&99.28&27438&83034 \\
11&pci\_bridge32&25653&23826&21914&45740&6535&1546&99.82&45742&230877 \\
12&pci\_spoci\_ctrl&1096&1523&1328&2851&196&85&99.04&2853&9864 \\
13&revx&7563&14575&14004&28579&1703&611&100&28581&68067 \\
14&sasc&889&602&514&1116&170&57&99.01&1118&8001 \\
15&simple\_spi&1135&930&794&1724&209&75&99.17&1726&10215 \\
16&spi&3890&4615&4301&8916&885&267&99.98&8918&35010 \\
17&sqrt32&2206&4216&3948&8164&442&170&99.85&8166&19854 \\
18&square&18080&30988&29880&60868&5640&1454&99.78&60870&162720 \\
19&ss\_pcm&604&313&257&570&97&31&98.59&572&5436 \\
20&systemcaes&11299&11100&10229&21329&2602&721&99.84&21331&101691 \\
21&systemcdes&3028&3312&3090&6402&627&195&99.49&6404&27252 \\
22&tv80&8177&11219&10368&21587&1672&578&99.73&21589&73593 \\
23&usb\_funct&16704&16054&14269&30323&2943&1057&99.77&30325&150336 \\
24&usb\_phy&599&538&460&998&140&37&97.47&1000&5391\\\hline
 \end{tabular}}
 \vspace{-0.3cm}
\end{table*}

%

%

 
The number of words~($S_D$) used determines the area of the mapping solution. To determine the effectiveness of the packing algorithm to pack blocks into words, we utilize the word utilization~($W_{Util}$) metric.  $W_{Util}$ is the percentage of total number of bits in $S_D$ words, that are used by the mapping solution. For the example MIG, out of 9 bits (3 words, each with 3 bits), 8 bits are used and therefore $W_{Util}$ is 88.8\%. The proposed packing algorithm achieves  more than 97\% utilization for all the benchmarks, when $w_D$ = 16, including 100\% utilization for the \textit{revx} benchmark. 
 Fig.~\ref{fig:util} shows that with increase in word length from 4 to 8, leads to considerable improvement in 
$W_{Util}$. However, the $W_{Util}$ is comparable for the word lengths, 8, 16 and 32, and approaches 100\%. This shows the effectiveness of delay-constrained mapping to pack the blocks into words.

 \section{Conclusion}\label{sec:conc}
\noindent In this work, we presented two approaches to the technology mapping problem for logic-in-memory computation using ReVAMP.
The area-constrained method allows high flexibility of addressing the need of mapping to a variety of crossbar dimensions while harnessing the available parallelism of the ReRAM crossbar array.
The delay-constrained method reduces the overall delay of mapping by using a multi-step approach that takes into account crossbar constraints while placing the operands. 
The proposed approach outperforms the state-of-the-art serial logic in memory approach using ReRAMs.

The synthesis approaches used in the technology mapping flow, such as partitioning algorithm used for LUT mapping, are not aware of the crossbar constraints. 
The LUT partitioning algorithm should ideally try to partition the graph, so that each of the ESOP expression corresponding to each LUT has roughly the same number of cubes, instead of solely minimizing the number of LUTs covering the graph. Also, the initial representation of the Boolean functions into AIG/MIG are not explicitly optimized w.r.t to the quality of the resulting mapping. 
We believe optimizing the synthesis algorithms w.r.t to the crossbar constraints, would allow further reduction in delay of mapping, when combined with the proposed technology mapping approaches.


%
\bibliographystyle{ieeetr}
\bibliography{reference}

\end{document}